\documentclass[twocolumn,aps,prl,superscriptaddress]{revtex4-2}

\usepackage{amsmath}
\usepackage{graphicx}
\usepackage{MnSymbol}
\usepackage{parskip}

\begin{document}

\title{Rubber wear: experiment and theory}

\author{B.N.J. Persson}
\affiliation{Peter Gr\"unberg Institute (PGI-1), Forschungszentrum J\"ulich, 52425, J\"ulich, Germany}
\affiliation{State Key Laboratory of Solid Lubrication, Lanzhou Institute of Chemical Physics, Chinese Academy of Sciences, 730000 Lanzhou, China}
\affiliation{MultiscaleConsulting, Wolfshovener str. 2, 52428 J\"ulich, Germany}
\author{R. Xu}
\affiliation{Peter Gr\"unberg Institute (PGI-1), Forschungszentrum J\"ulich, 52425, J\"ulich, Germany}
\affiliation{State Key Laboratory of Solid Lubrication, Lanzhou Institute of Chemical Physics, Chinese Academy of Sciences, 730000 Lanzhou, China}
\affiliation{MultiscaleConsulting, Wolfshovener str. 2, 52428 J\"ulich, Germany}
\author{N. Miyashita}
\affiliation{The Yokohama Rubber Company, 2-1 Oiwake, Hiratsuka, Kanagawa 254-8601, Japan}

\begin{abstract}
{\bf Abstract}: 
We study the wear rate (mass loss per unit sliding distance)
of a tire tread rubber compound sliding on concrete paver surfaces under dry and wet conditions, at different nominal contact pressures of $\sigma_0 = 0.12$, $0.29$, and $0.43 \ {\rm MPa}$, and sliding velocities ranging from $v= 1 \ {\rm \mu m/s}$ to $1 \ {\rm cm/s}$. We find that the 
wear rate is proportional to the normal force and remains independent of the sliding speed. 
Sliding in water and soapy water results in significantly lower wear rates compared to dry conditions. The experimental data are analyzed using a theory that predicts wear rates and wear particle size distributions consistent with the experimental observations.

\end{abstract}

\maketitle

\setcounter{page}{1}
\pagenumbering{arabic}




{\bf 1 Introduction}

Wear is the progressive loss of material from a solid body due to its contact and relative movement against a surface \cite{wear1,wear2,Rabi1, Rabi2, Moli1, Moli2,Roland}. Rubber wear is of great practical importance, e.g., tires and conveyor belts \cite{Wear, wear1, my}. Tire wear is the largest source of polymer (plastic) particles, and this source may increase with the increasing use of electric vehicles, as they are generally heavier than combustion engine vehicles \cite{particles}. Wear particles produced on road surfaces span a wide range of scales, from nanometers to millimeters, but the largest mass fraction consists of particles with diameters in the range of $1-1000 \ {\rm \mu m}$. Particles of this size typically result from the detachment of rubber fragments from surfaces through crack propagation. Smaller particles, particularly nanoparticles, may form due to stress corrosion, where bond-breaking barriers are lowered by interactions with foreign molecules, such as oxygen or ozone \cite{corr}.

There are several limiting cases of rubber wear, known as {\it fatigue wear}, {\it abrasive wear}, and {\it smearing wear}. When a rubber block slides on a rigid countersurface with ``smooth roughness" (to be defined in Sec. 7), the stress concentrations in asperity contact regions are relatively low, and many contacts with substrate asperities are needed to remove rubber particles. This results in fatigue failure rather than tensile failure, and the abrasion of rubber caused by this failure mode is called fatigue wear.

Abrasive wear occurs when a rubber block slides against surfaces with sharp asperities. In this case, stress concentrations generated by the sharp points of contact cut into the rubber, potentially reaching the material's limiting strength, leading to micro-cutting or scratching on the rubber surface. This process produces longitudinal scratches parallel to the sliding direction, known as score lines.

When rubber compounds are abraded under mild conditions, a sticky, gooey transfer layer often forms on the rubber and countersurface. The abrasion failure in this case is a type of degradation process referred to as smearing. This smearing is likely due to some form of rubber decomposition and may result from stress corrosion. It is worth noting that very small (e.g., nanoscale) particles tend to adhere to almost any surface, which can macroscopically appear as a sticky smear film \cite{RW1}.

Wear particles from tires on road surfaces often contain not only rubber but also road wear particles and dust (e.g., pollen or sand particles), making it challenging to compare wear studies on road surfaces with theoretical predictions. Wear particles may vary in size depending on how they are generated and collected. For instance, airborne wear particles tend to be smaller than those found on road surfaces. Additionally, wear particles produced in water tend to be smaller than those generated in dry conditions. This difference may result from the influence of water on wear processes, which is supported by the observation that the wear rate in water can be very different from that in dry conditions. Furthermore, in water, there is a lower probability of particle agglomeration compared to the dry state, where larger wear particles often consist of agglomerates of smaller particles.

Here, we focus on rubber wear particles produced under well-defined conditions in laboratory environments\cite{RW1}. An interesting study on this topic was presented in Ref. \cite{Wear}, where detailed results were provided for the wear of three tire tread compounds on three different sandpaper surfaces at three different humidity levels. Most wear particles were in the size range of $10-400 \ {\rm \mu m}$, but the wear rate varied significantly depending on the system. Specifically, the wear rate increased with increasing surface roughness, decreased with decreasing humidity, and was lower for two carbon black-filled compounds compared to a silica-filled compound.

Rubber crack propagation is crucial for understanding the origin of rubber wear. The crack or tearing energy $\gamma$ (usually denoted by $T$, but here we use $\gamma$ to avoid confusion with temperature) is defined as the energy per unit area required to separate surfaces at a crack tip\cite{Paris}. For rubber-like materials, $\gamma$ can be substantial, 
typically ranging from $\sim 10^2$ to $\sim 10^5 \ {\rm J/m^2}$, depending on the crack tip velocity and temperature. This should be compared to the crack energy for (brittle) crystalline solids, which is on the order of $\sim 1 \ {\rm J/m^2}$, even for solids with strong covalent bonds like diamonds. The large $\gamma$ in rubber-like materials arises partly from the energy required to stretch polymer chains at the crack tip before breaking the (strong) covalent bonds, and partly from viscoelastic energy dissipation in the region ahead of the moving crack tip.

\begin{figure}
\includegraphics[width=0.40\textwidth,angle=0.0]{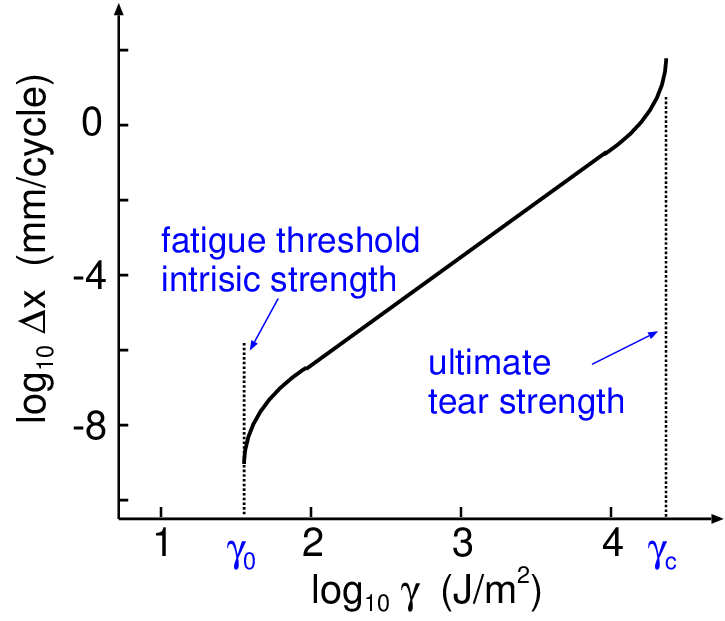}
\caption{\label{TearStrength.eps}
The crack growth length $\Delta x$ as a function of the tearing energy $\gamma$ (in a log-log scale). A crack is subjected to an oscillating strain with a typical frequency of $10 \ {\rm Hz}$. During each oscillation, the crack length increases by $\Delta x$, and the energy input corresponds to the tearing energy $\gamma \Delta A$, where $\Delta A = w \Delta x$ is the increase in crack surface area ($w$ the width of the crack surfaces).}
\end{figure}

In the literature, the crack energy $\gamma$ has been studied in detail in two cases: for crack tips moving at a constant velocity \cite{Gent} and for crack propagation in response to an oscillating strain \cite{Rivlin, NatRub, Ghosh}. Both sets of experiments yield similar results. In the case of an oscillating strain, we present schematic results for a rubber compound in Fig. \ref{TearStrength.eps}. The small value of $\Delta x$ indicates that, unless the applied strain (or stress) is large enough to bring $\gamma$ close to the ultimate tear strength, several stress cycles (resulting from the interaction with road asperities) may be needed to remove a particle from a rubber surface.

The tearing energy is usually measured in macroscopic rubber samples with a linear size of $\sim 1 \ {\rm cm}$, which may not be valid at the small length scales involved in rubber wear, where particles as small as $\sim 1 \ {\rm \mu m}$ may be removed. Specifically, the viscoelastic contribution to the tearing or crack energy may be reduced due to finite-size effects \cite{finite}. Additionally, during sliding, the asperity-induced deformation frequencies $\omega \approx v/r_0$ depend on the sliding speed $v$ and the size $r_0$ of the contact region, resulting in a broad range of frequency values, while the tearing energy is usually measured at a fixed frequency.

In this paper, we study the wear rate of a tire rubber tread compound (carbon black-filled natural rubber used for bus and truck tires) sliding on concrete surfaces under dry and wet conditions, at different nominal contact pressures ($\sigma_0 = 0.12$, $0.29$, and $0.43 \ {\rm MPa}$) and sliding speeds ranging from $1 \ {\rm \mu m/s}$ to $1 \ {\rm cm/s}$. We find that the wear rate is proportional to the normal force and independent of the sliding speed. Sliding in water and soapy water results in wear rates that are much lower than in dry conditions. The experimental data are analyzed using a new theoretical approach that predicts wear rates and wear particle sizes consistent with the experimental observations.

\begin{figure}
\includegraphics[width=0.47\textwidth,angle=0.0]{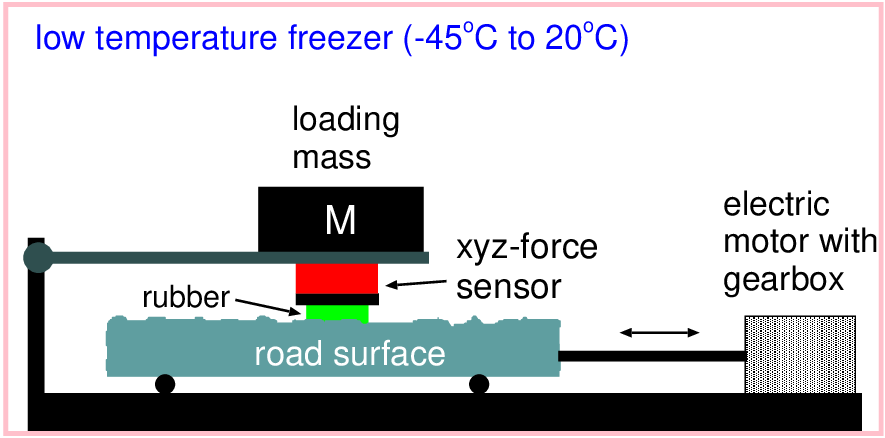}
\caption{\label{LowTemperaturePic.eps}
Schematic picture of low-temperature friction instrument allowing for linear reciprocal motion.
}
\end{figure}

\vskip 0.3cm
{\bf 2 Experimental methods}

We measured the friction coefficient and wear rate using the experimental setup shown in Fig. \ref{LowTemperaturePic.eps}. A rectangular rubber block, $3 \ {\rm cm}$ in length along the sliding direction and $7 \ {\rm cm}$ in the perpendicular direction, is glued into the milling groove of the sample holder, which is attached to the force cell (red box in the figure). The rubber specimen can move vertically with the carriage to adapt to the substrate profile. The normal load can be changed by adding additional steel weights on top of the force cell. The substrate sample is mounted on the machine table, which is moved in a translational manner by a servo drive through a gearbox. Here, we control the relative velocity between the rubber specimen and the substrate, while the force cell records data on the normal force and friction.

To study the velocity and pressure dependency of the friction coefficient and wear rate, we slide the rubber sample on the surfaces of concrete blocks (concrete pavers) at different velocities and normal forces (loads). Each sliding cycle consists of $20 \ {\rm cm}$ forward and $20 \ {\rm cm}$ backward motion. The wear rate is determined from the mass change (the difference in the mass of rubber blocks before and after sliding) using a high precision balance (Mettler Toledo analytical balance, model MS104TS/00) with a sensitivity of $0.1 \ {\rm mg}$. Except for the run-in, we replace the concrete block after each sliding cycle so that each measurement is conducted on a fresh concrete surface.

\vskip 0.3cm
{\bf 3 Experimental results}

All experiments have been performed using a bus/truck tread compound consisting of 
natural rubber with carbon black (55weight\% natural rubber and 30weight\% 
carbon black and less than 1weight\% polybutadiene rubber). To reduce the influence of frictional heating, all tests were performed at low sliding speeds. We begin by describing the run-in process of the rubber surfaces, followed by a detailed study of the velocity and load dependency of friction and wear rate. Additionally, we present results for sliding in water and soapy water.

\begin{figure}
\includegraphics[width=0.47\textwidth,angle=0.0]{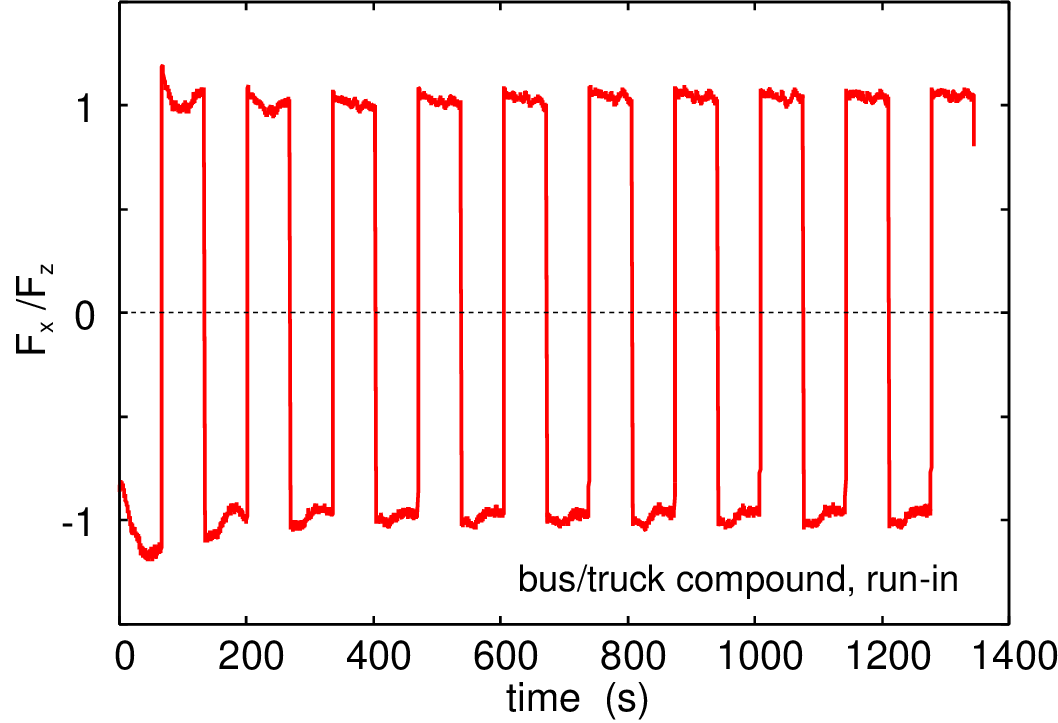}
\caption{
	The ratio $F_x/F_z$ during run-in of a truck/bus tread compound 
	sliding on a concrete surface. Each run-in sliding cycle consists of $20 \ {\rm cm}$ forward and $20 \ {\rm cm}$
	backward motion. The sliding speed $v=3 \ {\rm mm/s}$, the temperature $T=23^\circ {\rm C}$ and the nominal
	contact pressure $p=0.1 \ {\rm MPa}$. 
}
\label{1time.2FxOverFz.BusRacing2a.eps}
\end{figure}

\begin{figure}
\includegraphics[width=0.47\textwidth,angle=0.0]{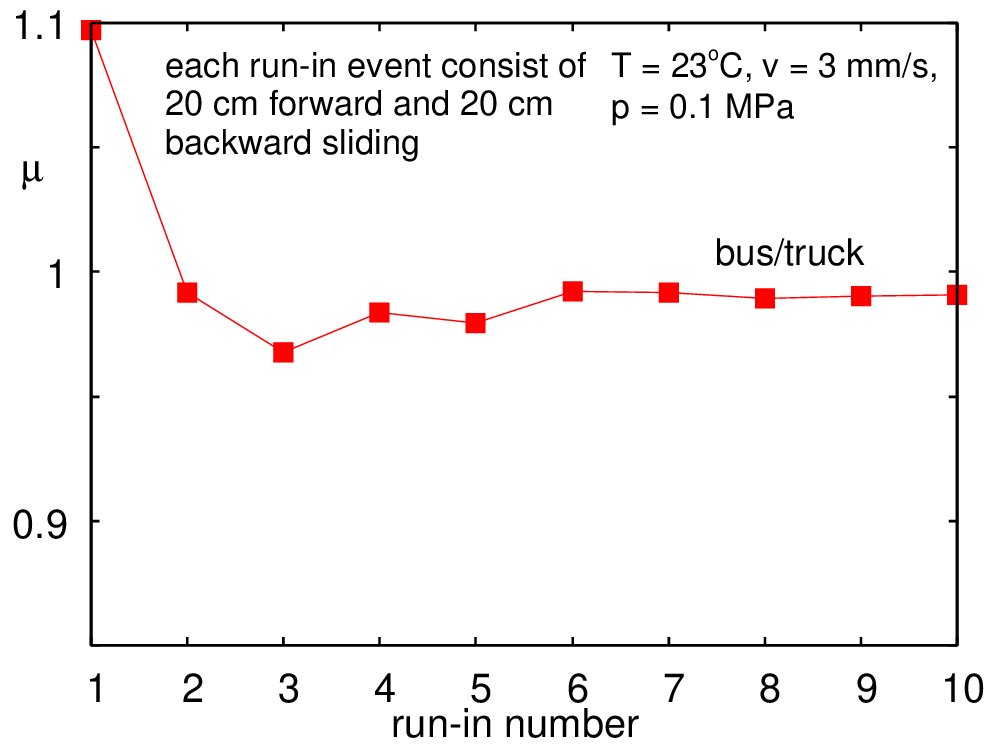}
\caption{
	The friction coefficient during run-in of a truck/bus tread compound 
	sliding on a concrete surface. Each run-in sliding cycle consists of $20 \ {\rm cm}$ forward and $20 \ {\rm cm}$
	backward motion. The sliding speed $v=3 \ {\rm mm/s}$, the temperature $T=23^\circ {\rm C}$ and the nominal
	contact pressure $p=0.1 \ {\rm MPa}$.
}
\label{1number.2mu.blueRacingRedBus2.eps}
\end{figure}

\vskip 0.1cm
{\bf Run-in}

Fig. \ref{1time.2FxOverFz.BusRacing2a.eps} shows 
	the ratio $F_x/F_z$ between the tangential and normal force during run-in 
	on a concrete surface. Each run-in sliding cycle consists of $20 \ {\rm cm}$ forward and $20 \ {\rm cm}$
	backward motion. 
The sliding speed $v=3 \ {\rm mm/s}$, the temperature $T=23^\circ {\rm C}$ and the nominal
	contact pressure $p=0.1 \ {\rm MPa}$. 

Fig. \ref{1number.2mu.blueRacingRedBus2.eps} shows the friction coefficient $\mu$, here defined as the 
average of $|F_x|/F_z$ during one sliding cycle, as a function of the number of run-in cycles.
The friction coefficient stabilized at $\mu \approx 1$ after $\sim 6$ sliding cycles.

\begin{figure}
\includegraphics[width=0.47\textwidth,angle=0.0]{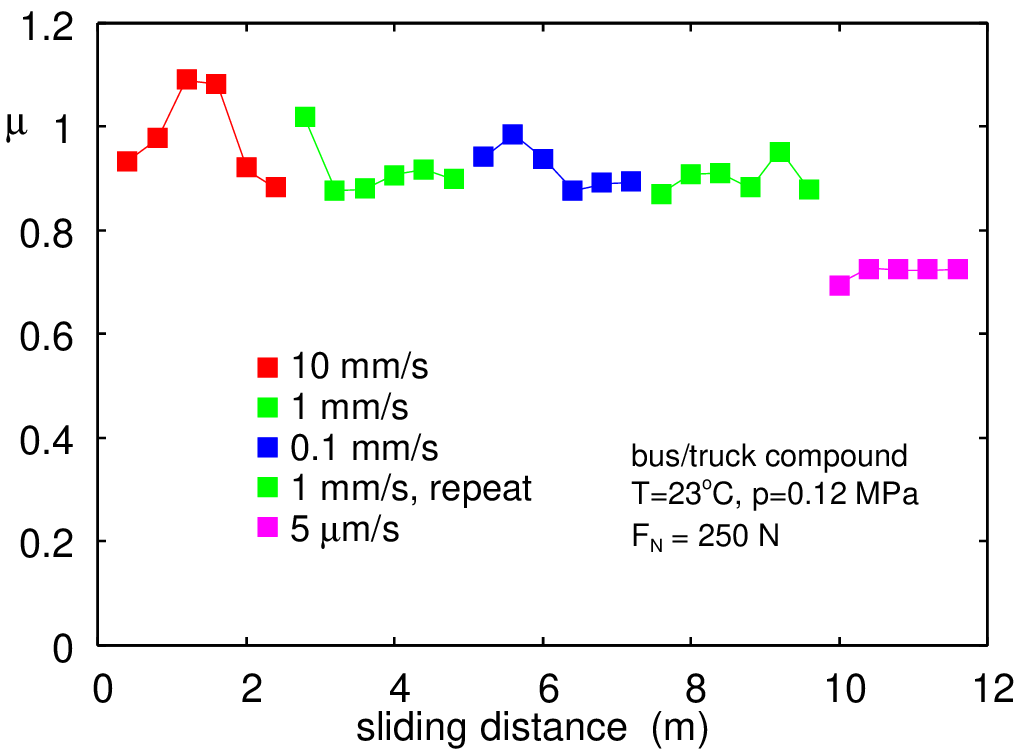}
\caption{
	The average friction coefficient as a function of total sliding distance for several 
sliding speeds.	Each data point corresponds to a sliding cycle ($20 \ {\rm cm}$ forwards and $20 \ {\rm cm}$ backward motion) on fresh concrete surfaces (new concrete block for each sliding cycle). 
}
\label{1slidingdistance.2mu.all.velocities.with.5mum.eps}
\end{figure}

\begin{figure}
\includegraphics[width=0.47\textwidth,angle=0.0]{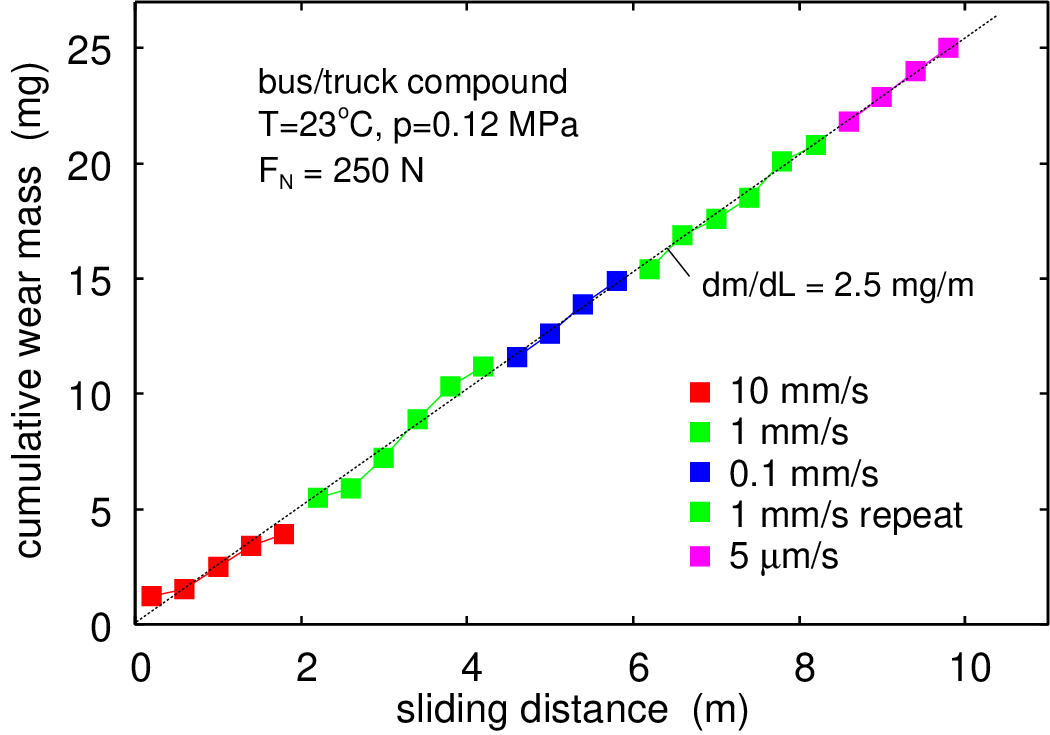}
\caption{
	The cumulative rubber wear in ${\rm mg}$ during 5 sliding cycles ($20 \ {\rm cm}$ forwards and $20 \ {\rm cm}$ backward motion)
	on fresh concrete surfaces (new concrete block for each sliding cycle) as a function of the total sliding distance.
	The red, green, and blue symbols are for the sliding speeds $v=10$, $1$ and $0.1 \ {\rm mm/s}$.
}
\label{1l.3dm.for.all.velocities.eps}
\end{figure}

\begin{figure}
\includegraphics[width=0.47\textwidth,angle=0.0]{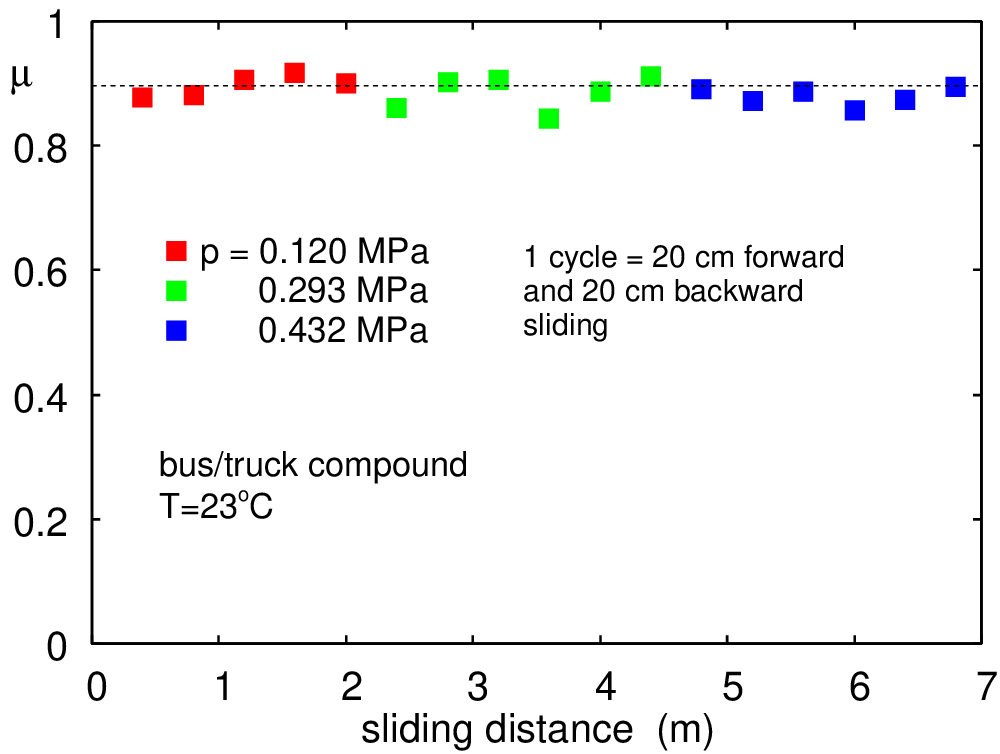}
\caption{
	The friction coefficient as a function of the sliding distance. 
	Each data point is the result of
	a sliding cycle ($20 \ {\rm cm}$ forwards and $20 \ {\rm cm}$ backward motion)
	on fresh concrete surfaces (new concrete block for each sliding cycle).
	The red, green, and blue symbols are for the nominal contact pressures $p=0.120$, $0.293$ and $0.432 \ {\rm MPa}$.
	For the sliding speed $v=1 \ {\rm mm/s}$.
}
\label{1slidingDistance.2mu.all.loads.eps}
\end{figure}

\begin{figure}
\includegraphics[width=0.47\textwidth,angle=0.0]{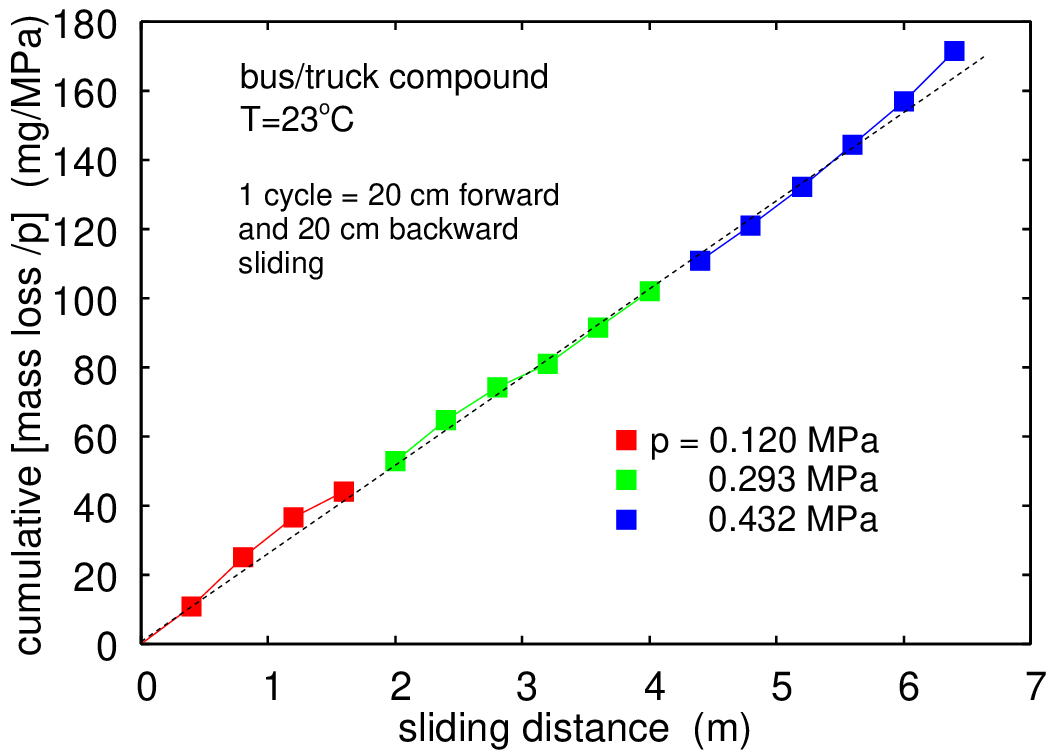}
\caption{
	The cumulative wear mass $m$ over pressure $p$ in ${\rm mg/MPa}$ 
        as a function of the sliding distance.
	Each data point is the result of
	a sliding cycle ($20 \ {\rm cm}$ forwards and $20 \ {\rm cm}$ backward motion)
	on fresh concrete surfaces (new concrete block for each sliding cycle).
	The red, green, and blue symbols are for the nominal contact pressures $p=0.120$, $0.293$ and $0.432 \ {\rm MPa}$.
	For the sliding speeds $v=1 \ {\rm mm/s}$.
}
\label{1distance.2cumulativeWear.all.pressures.eps}
\end{figure}

\vskip 0.1cm
{\bf Friction and wear on dry concrete surfaces}

Fig. \ref{1slidingdistance.2mu.all.velocities.with.5mum.eps} shows the friction coefficient as a function of total sliding distance for several sliding speeds. Each data point corresponds to the friction averaged over a sliding cycle ($20 \ {\rm cm}$ forward and $20 \ {\rm cm}$ backward motion) on a fresh concrete surface (i.e., a new concrete block for each sliding cycle). Note that the friction coefficient remains nearly constant at $\sim 1$ as the sliding speed varies between $0.1 \ {\rm mm/s}$ and $10 \ {\rm mm/s}$ but drops to $\sim 0.75$ at a sliding speed of $5 \ {\rm \mu m/s}$.

Fig. \ref{1l.3dm.for.all.velocities.eps} shows the cumulative rubber wear in ${\rm mg}$ as a function of total sliding distance for the same sliding cycles as in Fig. \ref{1slidingdistance.2mu.all.velocities.with.5mum.eps}. Note that within the noise of the measured data and the tested sliding speed range, the wear rate is independent of sliding speed and $\approx 2.5 \ {\rm mg/m}$.

Fig. \ref{1slidingDistance.2mu.all.loads.eps} shows the friction coefficient as a function of sliding distance for nominal contact pressures of $p=0.120$, $0.293$, and $0.432 \ {\rm MPa}$ at a sliding speed of $v=1 \ {\rm mm/s}$. Note that within the studied pressure range, the friction coefficient is independent of the nominal contact pressure.

Fig. \ref{1distance.2cumulativeWear.all.pressures.eps} shows the cumulative [rubber wear mass over pressure] ($m/p$) in ${\rm mg/MPa}$ as a function of sliding distance for the same sliding cycles as in Fig. \ref{1slidingDistance.2mu.all.loads.eps}. The red, green, and blue symbols represent contact pressures of $p=0.120$, $0.293$, and $0.432 \ {\rm MPa}$, respectively. Note that within the studied pressure range, the wear rate is proportional to the nominal contact pressure.

\begin{figure}
\includegraphics[width=0.47\textwidth,angle=0.0]{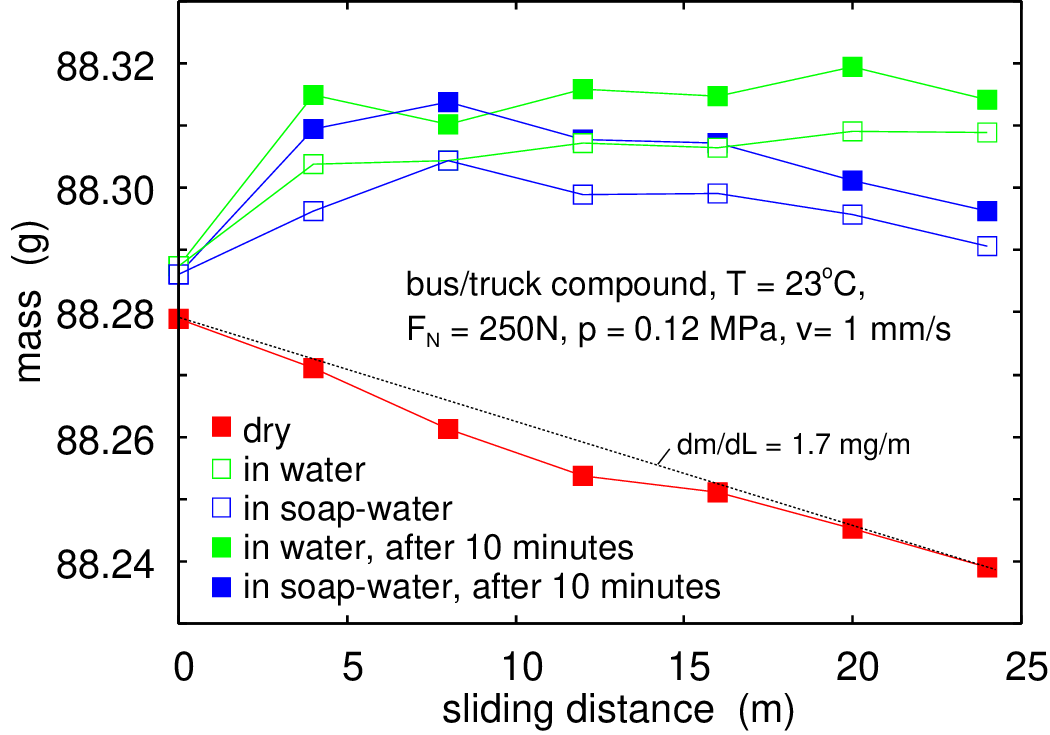}
\caption{
The mass of the sample (rubber plus aluminum plate) as a function of sliding distance for sliding in dry conditions (red), distilled water (green), and soapy water (blue). Each data point represents the result of 10 sliding cycles (each consisting of $20 \ {\rm cm}$ forward and $20 \ {\rm cm}$ backward motion), totaling $4 \ {\rm m}$ of sliding on fresh concrete surfaces. After each sliding cycle in water or soapy water, the rubber was dried with a paper towel, and the mass was measured immediately (open squares) and a second time after being left for 10 minutes in ambient conditions (filled squares). The dry rubber wear experiment was conducted approximately one month after the soapy water experiment, indicating some mass loss during the waiting period.
}

\label{1distance.2wearrate.redDRY.greenWATER.blueSOAPWATER.eps}
\end{figure}

\vskip 0.1cm
{\bf Friction on wet concrete surfaces}

Fig. \ref{1distance.2wearrate.redDRY.greenWATER.blueSOAPWATER.eps} shows the mass of the sample (rubber plus aluminum plate) as a function of sliding distance for sliding in dry conditions (red), distilled water (green), and soapy water (blue). Each data point represents the result of 10 sliding cycles (each consisting of $20 \ {\rm cm}$ forward and $20 \ {\rm cm}$ backward motion), totaling $4 \ {\rm m}$ of sliding on fresh concrete surfaces. After each sliding cycle in water or soapy water, the rubber was dried with a paper towel, and the mass was measured immediately (open squares) and a second time after being left for 10 minutes in ambient conditions (filled squares). Note that the mass of the block at the start of sliding in water or soapy water is nearly the same as at the start of sliding, indicating little or no mass loss for wet conditions. The dry rubber wear experiment was conducted approximately one month after the wet conditions experiment, suggesting some mass loss occurred during the waiting period.

Fig. \ref{1distance.2wearrate.redDRY.greenWATER.blueSOAPWATER.eps} shows that the wear rate in water and soapy water is much lower than in dry conditions, and much longer sliding distances are needed to determine the wear 
rate (if exist) in water and soapy water.

\begin{figure}
\includegraphics[width=0.47\textwidth,angle=0.0]{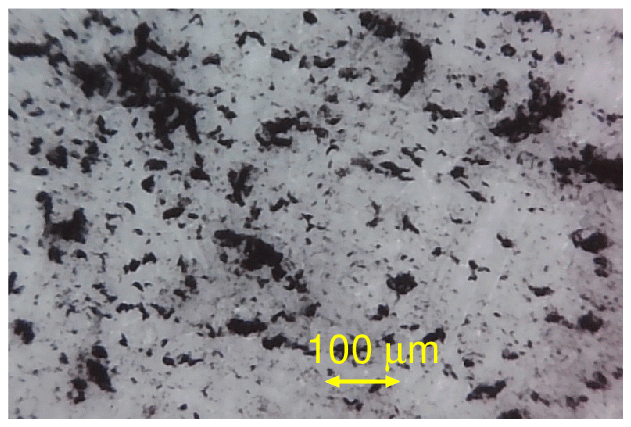}
\caption{
Rubber wear particles collected using adhesive tape. The particles were removed from the rubber surface with adhesive tape after sliding the rubber block onto the concrete surface. The largest wear particles may be agglomerates of smaller particles.
}
\label{RubberParticle2.ps}
\end{figure}

\vskip 0.1cm
{\bf Rubber wear particle sizes}

We studied the size of the rubber wear particles using an optical microscope. The particles were picked up from the rubber surface using adhesive tape. A rubber block was first slid on a clean, dry concrete surface, after which adhesive tape was pressed onto the rubber block surface. Fig. \ref{RubberParticle2.ps} shows the rubber particles collected by the adhesive tape. The typical size of the rubber particles is about $1-100 \ {\rm \mu m}$, which aligns with findings from other studies. However, it is possible that some of the larger wear particles are agglomerates of smaller particles.

\vskip 0.3cm
{\bf 4 Theory of rubber wear}

\begin{figure}
\includegraphics[width=0.25\textwidth,angle=0.0]{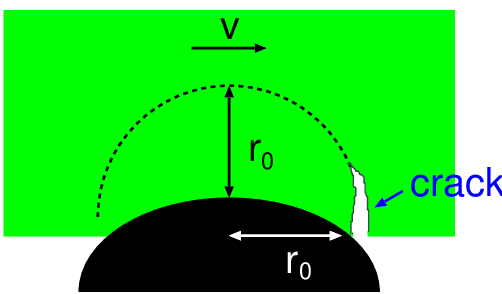}
\caption{\label{WearParticlea.eps}
A rubber block sliding in contact with a road asperity. The sliding speed $v$ and the radius of the contact region $r_0$ are indicated. The deformation field extends into the rubber a similar distance as it extends laterally.
}
\end{figure}

Cracks at a rubber surface 
can be induced by both the compressive and the tangential stress acting on the surface,
but particle removal is caused mainly by the tangential stress.
Let $\tau = \tau (\zeta_r)$ be the effective 
shear stress acting in an asperity contact region with radius $r_0$.
The magnification $\zeta_r$ is determined by the radius of the contact region,
$q_r = \pi /r_0$,  $\zeta_r = q_r /q_0$.
The elastic energy stored in the deformed asperity contact is (see Fig. \ref{WearParticlea.eps})
$$U_{\rm el} \approx {\tau^2  \over  E^*} r_0^3$$
where the effective modulus $E^* = E/(1-\nu^2)$ (we assume that the substrate is rigid).
More accurately, assume that the shear stress acts uniformly within a circular region
with radius $r_0$. The center of the circular region will displace a distance $u$ given by
$k u = F$ where $F=\tau \pi r_0^2$ is the force and $k \approx (\pi/2) E^* r_0$ the spring constant.
This gives the elastic energy
$$U_{\rm el} = {1\over 2} k u^2 = {F^2 \over 2 k} = 
{(\pi r_0^2 \tau)^2\over \pi E^* r_0} = \pi {\tau^2 \over E^*} r_0^3\eqno(1)$$
In order for the shear stress to 
remove a particle of size $r_0$ the stored elastic energy must be larger than the
fracture (crack) energy which is of order
$$U_{\rm cr} \approx \gamma 2 \pi r_0^2 \eqno(2)$$
where $\gamma$ is the energy 
per unit surface area to break the bonds at the crack tip. If $U_{\rm el}>U_{\rm cr}$
the elastic energy is large enough to propagate a crack and
remove a particle of the linear size $r_0$. Thus for a particle to be 
removed we must have $\tau > \tau_{\rm c}$ where
$$\tau_{\rm c} = \beta \left ({2 E^* \gamma \over r_0 }\right )^{1/2}\eqno(3)$$
where $\beta$ is a number of order unity which takes into account that the wear particles in general
are not spherical cups as assumed above.

We will first assume that whenever $U_{\rm el}>U_{\rm cr}$ is obeyed a wear particle of 
size $r_0$ is removed. This is not the case in rubber wear where in general many 
contacts between a rubber crack and road asperities are needed to remove a rubber 
particle, and we will later ``correct'' this fact.

In what follows we will treat the rubber surface as 
smooth and assume only roughness on the road surface.
In most tire applications this is a good 
approximation as even for a worn rubber surface the road surface
has higher roughness than the rubber surface.
We will denote a road asperity where the shear stress is high enough to remove a particle
of size $r_0$ as a {\it wear-asperity} and the corresponding contact region as the {\it wear-contact region}.
 
If we assume that during sliding the effective 
shear stress $\tau$ is proportional to the normal stress $\sigma$, $\tau = \mu \sigma$
we get that particles will get removed only if the contact stress $\sigma > \sigma_{\rm c} (\zeta)$ where
$$\sigma_{\rm c} = {\beta \over \mu} \left ({2 E^* \gamma \over r_0 }\right )^{1/2}\eqno(4)$$ 

For randomly rough surfaces the probability distribution of contact stress equals (see Appendix A):
$$P(\sigma,\zeta) = {1\over (4\pi G)^{1/2}} \left (e^{-(\sigma-\sigma_0)^2/4G}- e^{-(\sigma+\sigma_0)^2/4G} \right )\eqno(5)$$
where $\sigma_0$ is the nominal (applied) pressure and where
$$G ={\pi \over 4} (E^*)^2 \int_{q_0}^{\zeta q_0} dq \ q^3 C(q)\eqno(6)$$
where $C(q)$ is the surface roughness power spectrum.
When the interface is studied at the magnification $\zeta$ 
the area $A=A_{\rm wear}(\zeta)$ where the shear stress is high enough to remove particles is given by
$${A_{\rm wear} (\zeta) \over A_0} = \int_{\sigma_{\rm c} (\zeta)}^\infty d\sigma \ P(\sigma,\zeta)\eqno(7)$$

If we assume that the wear-asperity contact regions are circular we can write
the wear area as $A_{\rm wear} = N \pi r_0^2$. If every contact with a wear asperity 
remove a rubber particle (severe wear)
the wear volume would consist of $N$ cylinders of length $L$
and cross-section area $\pi r_0^2/2$. Hence in this limit, the wear volume would be
$V= N L \pi r_0^2/2$ or
$${V\over L A_0} = {N \pi r_0^2 \over 2 A_0} = {1\over 2} {A_{\rm wear} \over A_0}\eqno(8)$$
Using (7) this gives
$${V (\zeta) \over L} = {A_0 \over 2} \int_{\sigma_{\rm c} (\zeta)}^\infty d\sigma \ P(\sigma,\zeta)\eqno(9)$$
Thus the wear volume per unit sliding length is proportional to the nominal surface area, as expected
when the nominal contact pressure is constant.

\begin{figure}
\includegraphics[width=0.25\textwidth,angle=0.0]{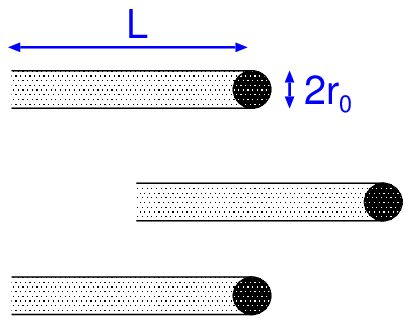}
\caption{\label{wearline.eps}
When the rubber block is slid on the concrete surface the road wear-asperities
remove rubber particles along lines.
A line of length $L$ gives a wear volume $L \pi r_0^2/2$. This picture is valid
in the limit of severe wear where every contact between the rubber and a road 
wear-asperity results in a wear particle. For mild wear, many contacts between the rubber and the
road wear-asperities are necessary to remove a wear particle. 
}
\end{figure}

When the crack (or fracture) energy $\gamma$ is independent of the crack tip speed one expects a
wear particle to form whenever the stored elastic energy $U_{\rm el}$ is larger than the fracture energy 
$U_{\rm cr} \approx \gamma \pi r_0^2$.
However, for rubber materials the crack energy increases strongly with the crack tip speed $v_{\rm cr}$.
In this case, if the stored elastic energy is close to the $\gamma (v_{\rm cr}=0) \pi r_0^2$, 
the crack tip will move only a very short distance $\Delta x << r_0$ 
during the time it takes for a wear asperity to pass the crack.
In this case very many interactions between the crack and wear asperities are needed to remove a wear
particle of size $r_0$. 
Here we will assume that if the elastic energy $U_{\rm el} >> U_{\rm el0}$, where
$U_{\rm el0} = \gamma (v_{\rm cr}=0) \pi r_0^2$,  the crack tip will move with the
velocity $v_{\rm cr}$ such that $U_{\rm el} = \gamma (v_{\rm cr}) \pi r_0^2$. 
The crack-tip displacement $\Delta x = v_{\rm cr} \Delta t$
during the time $\Delta t$ the crack interact with an asperity. Since $\Delta t \approx r_0/v$, where $v$ is
the sliding speed, this argument would indicate that the displacement $\Delta x$ depends on the
sliding speed but we have found that this is not the case (see Sec. 3). One explanation for this could
be that the rubber at the interface does not slip uniformly with the applied (or average) speed but
performs stick-slip motion at the asperity level involving slip speeds independent of the driving speed. 
In what follows we will assume that $\Delta x$ does not depend on the sliding speed but we will consider it a
function only of the tearing energy $\gamma$. This assumption cannot be exact and needs further study.

The number of contacts needed to remove a particle $N_{\rm cont} \approx r_0/\Delta x$ 
depend on the crack energy $\gamma$ but could be a large number ($10^3$ or more) 
if the macroscopic relation between the tear-energy 
$\gamma$ and $\Delta x$ would hold also at the length scale of the
wear particles (see Sec. 8). However, for the small $r_0$ of interest here $N_{\rm cont}$ may be smaller
because of a reduction in the viscoelastic dissipation in front of the crack tip
(finite-size effect).

We now take into account that a crack moves only a short distance $\Delta x$ in 
each contact with a wear-asperity. Assume that the wear-asperities are
randomly distributed in the nominal contact area $A_0$.
If a wear particle is removed in a single contact with a wear-asperity (as assumed in deriving (8)) then
the sliding distance $L^*$ needed to remove one layer of particles (thickness $r_0$) with volume $A_0 r_0$ 
is given by (8) with $V=A_0 r_0$:
$${r_0 \over L^*} = {1\over 2} {A_{\rm wear} \over A_0}$$
or
$$L^* = 2 r_0 {A_0 \over A_{\rm wear}}\eqno(10)$$
After the sliding distance $L^*$ every point 
on the rubber surface has been in contact with a wear-asperity
as this is the condition for removing a particle when $N_{\rm cont}=1$. 
After the sliding distance $N_{\rm cont} L^*$
every point on the rubber surface has been in contact with a wear-asperity $N_{\rm cont}$ times.
Hence if $N_{\rm cont}$ contacts is needed to remove a particle then
the sliding distance to remove one layer of particles will be $N_{\rm cont} L^*$. This gives
$${V\over A_0 L} = {V \over A_0 N_{\rm cont} L^*} = {r_0 \over N_{\rm cont} L^*}$$
$$={1\over 2 N_{\rm cont}} {A_{\rm wear} \over A_0}$$
We write
$$N_{\rm cont} = 1+r_0/\Delta x\eqno(11)$$
This equation interpolates between  
$N_{\rm cont} = r_0/\Delta x$ as $\Delta x/r_0 \rightarrow 0$
and $N_{\rm cont} = 1$ as $\Delta x/r_0 \rightarrow \infty$. 
Using this we get
$${V\over A_0 L} =
{1\over 1+r_0/\Delta x} {A_{\rm wear} \over 2 A_0}
={1\over 2} \int_{\sigma_{\rm c} (\zeta)}^\infty d\sigma \ {P(\sigma,\zeta) \over 1+r_0/\Delta x}\eqno(12)$$
In this equation $\Delta x(\gamma)$ is a function of the tearing energy $\gamma$.
For rubber-like materials, $\gamma$ is not a fixed number but takes a range of 
values $\gamma_0 < \gamma < \gamma_{\rm c}$. We can take this into account in (12) by considering
$\gamma $ as a function of $\sigma$ and $\zeta$ given by (4) or
$$\gamma = \left ({\mu \sigma \over \beta }\right )^2 {r_0 (\zeta) \over 2 E^*}\eqno(13)$$
Denoting $\Delta x(\gamma)$ by $\Delta x(\sigma,\zeta)$ we get
$${V\over A_0 L} = {1\over 2} 
\int_{\sigma_{\rm c} (\zeta)}^\infty d\sigma \ {P(\sigma,\zeta) \over 1+r_0(\zeta)/\Delta x(\sigma,\zeta)}\eqno(14)$$

One layer of wear particles corresponds to $N^* = A_0/\pi r_0^2$ wear particles. Hence the number of wear particles
removed per unit sliding distance is
$${N \over A_0 L} = {N^* \over N_{\rm cont} A_0 L^*} = {1\over \pi r_0^2} {A_{\rm wear}\over A_0}
{1\over 2 r_0 N_{\rm cont}} \eqno(15)$$
or using (11),
$$ {N \over A_0 L} = {1 \over 2 \pi r_0^3 }{1\over 1+r_0/\Delta x}  {A_{\rm wear} \over A_0}\eqno(16)$$
or when $\Delta x$ depends on $\gamma$:
$${N \over A_0 L} = {1 \over 2 \pi r_0^3 (\zeta) } \int_{\sigma_{\rm c} (\zeta)}^\infty d\sigma \ 
{P(\sigma,\zeta) \over 1+r_0(\zeta)/ \Delta x(\sigma,\zeta)}\eqno(17)$$

\begin{figure}
\includegraphics[width=0.35\textwidth,angle=0.0]{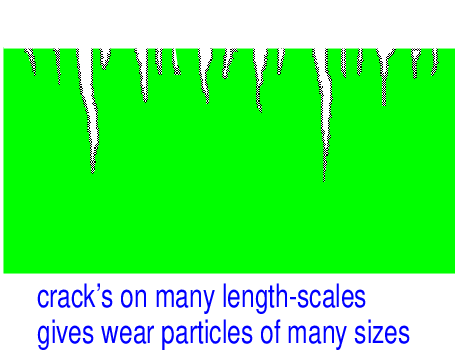}
\caption{\label{SurfaceCrackMany1.eps}
Cracks at a rubber surface form at many different length scales. Big road asperities generate ``long'' cracks and large wear particles. A big asperity has smaller asperities on top of it generating shorter cracks and smaller wear particles.
}
\end{figure}

\begin{figure}
\includegraphics[width=0.45\textwidth,angle=0.0]{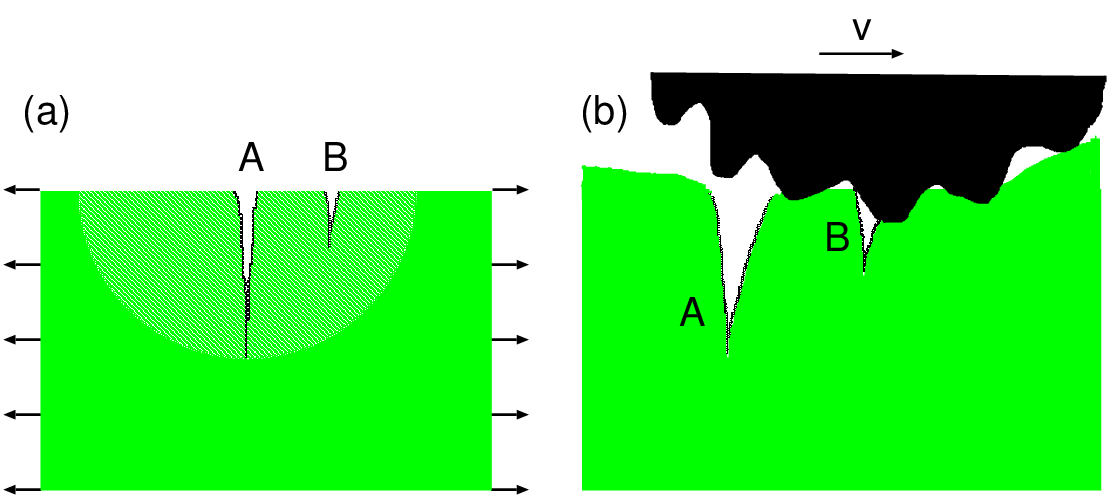}
\caption{\label{TwoCracksInteract.eps}
(a) A rubber sheet stretched parallel to the surface. The long crack A reduces the
tensile stress and the elastic energy density
in a hemispherical volume element with a radius of order the length of the
crack. This reduces the driving force for a smaller crack B in its vicinity (crack tip shielding).
(b) The contact region between a hard asperity and a rubber surface during sliding is
always under shear stress as long as slip occurs, and the stress and elastic deformation energy density 
at the small crack, B is only weakly influenced by the big crack A.
}
\end{figure}

The theory above gives the wear volume assuming that particles of a given size (radius $r_0$) are generated.
These are the (smallest) particles observed at the magnification $\zeta = q_r/q_0 = \pi / q_0 r_0$.
To get the total wear volume we need to sum up the volume of the wear particles from all length scales, 
which can be observed as we increase the magnification (see Fig. \ref{SurfaceCrackMany1.eps}). 
In order not to count particles of similar size
twice we will increase the magnification in steps of a 
factors of $\sim 2$ and write $\zeta = 2^n = \zeta_n$ where $n=0,1,..,n_1$ and $2^{n_1} q_0 = q_1$. 
We will refer to the interval from $\zeta = 2^n$ to $2^{n+1}$ as a 2-interval.
Using that
$$\sum_{n=0}^{n_1} f_n \approx \int_0^{n_1} dn \ f_n = 
{1\over {\rm ln}2} \int_1^{\zeta_1} d\zeta \ {1\over \zeta} f(\zeta)$$ 
we can write the total wear volume when $\Delta x$ is constant as
$${V\over A_0 L} \approx  {1\over 2}
\sum_{n=0}^{n_1} {1\over 1+r_0(\zeta_n)/\Delta x} {A_{\rm wear} (\zeta_n) \over A_0}$$
$$\approx {1\over 2 {\rm ln}2 }  \int_1^{\zeta_1} d\zeta \ {1\over \zeta} {1\over 1+r_0(\zeta)/\Delta x} 
{A_{\rm wear} (\zeta) \over A_0}\eqno(18)$$ 
Using that $\zeta r_0 = \pi/q_0$ this gives
$${V\over A_0 L} \approx  
{1\over 2 {\rm ln}2 }  \int_1^{\zeta_1} d\zeta \ {1\over \zeta+\pi/q_0\Delta x} {A_{\rm wear} (\zeta) \over A_0}$$
$$={1\over 2 {\rm ln}2 } \int_{q_0}^{q_1} dq  \ {1\over q+\pi/\Delta x} {A_{\rm wear} (q) \over A_0}\eqno(19)$$
When $\Delta x$ depends on $\gamma$ we get
$${V\over A_0 L} ={1\over 2 {\rm ln}2 } \int_{q_0}^{q_1} dq  
\int_{\sigma_{\rm c} (\zeta)}^\infty d\sigma \ 
{P(\sigma,\zeta) \over q+\pi/\Delta x(\sigma,\zeta)}\eqno(20)$$
where $\zeta = q/q_0$. It is convenient to write $q=q_0 e^\xi$ so that $dq=q d\xi$ and
$${V\over A_0 L} ={1\over 2 {\rm ln}2 } \int_{0}^{\xi_1} d\xi  
\int_{\sigma_{\rm c} (\zeta)}^\infty d\sigma \ 
{P(\sigma,\zeta) \over 1+r_0(\zeta)/\Delta x(\sigma,\zeta)}\eqno(21)$$
where $\xi_1 = {\rm ln}(q_1/q_0)$.

The distribution of particles of different sizes is given by (16) [or (17)]. Thus the number of
particles with radius $r_0$ between $(\pi / q_0) 2^{-n-1/2}$ and $(\pi / q_0) 2^{-n+1/2}$ is
$${N_n \over A_0 L} \approx {1 \over 2 \pi r_0^3 (\zeta_n) [1+r_0 (\zeta_n)/\Delta x]} {A_{\rm wear} (\zeta_n) \over A_0}\eqno(22)$$
or when $\Delta x$ depends on $\gamma$
$${N_n \over A_0 L} \approx {1 \over 2 \pi r_0^3 (\zeta_n) } 
\int_{\sigma_{\rm c} (\zeta_n)}^\infty d\sigma \ 
{P(\sigma,\zeta_n) \over 1+r_0(\zeta_n)/\Delta x(\sigma,\zeta_n)}\eqno(23)$$

The theory presented above assumes that all length scales contribute independently to the wear rate.
This cannot be strictly true since a long crack, which would result in a large wear
particle, will change the stress field in its vicinity out to a distance of order the length of the crack.
This effect, known as crack shielding (to be discussed later), reduces the ability for smaller cracks to grow in the neighborhood of longer cracks. However, crack tip shielding is much weaker for sliding contacts as compared to rubber strips elongated by uniform far-field stress (see Fig. \ref{TwoCracksInteract.eps}). 

Note that if $r_0/\Delta x$ is large a long run-in distance would be needed before the wear reaches a steady state.
This is particularly true if the nominal contact pressure is small where the distance between the wear asperity
contact regions may be large. However, since the contact regions within the macroasperity contacts are densely distributed
and independent of the nominal contact pressure, there may, in some cases be enough wear asperity contact regions within the
macroasperity contact regions to reach the $N_{\rm cont}$ needed for wear particle formation 
even over a short sliding distance.

\begin{figure}
\includegraphics[width=0.47\textwidth,angle=0.0]{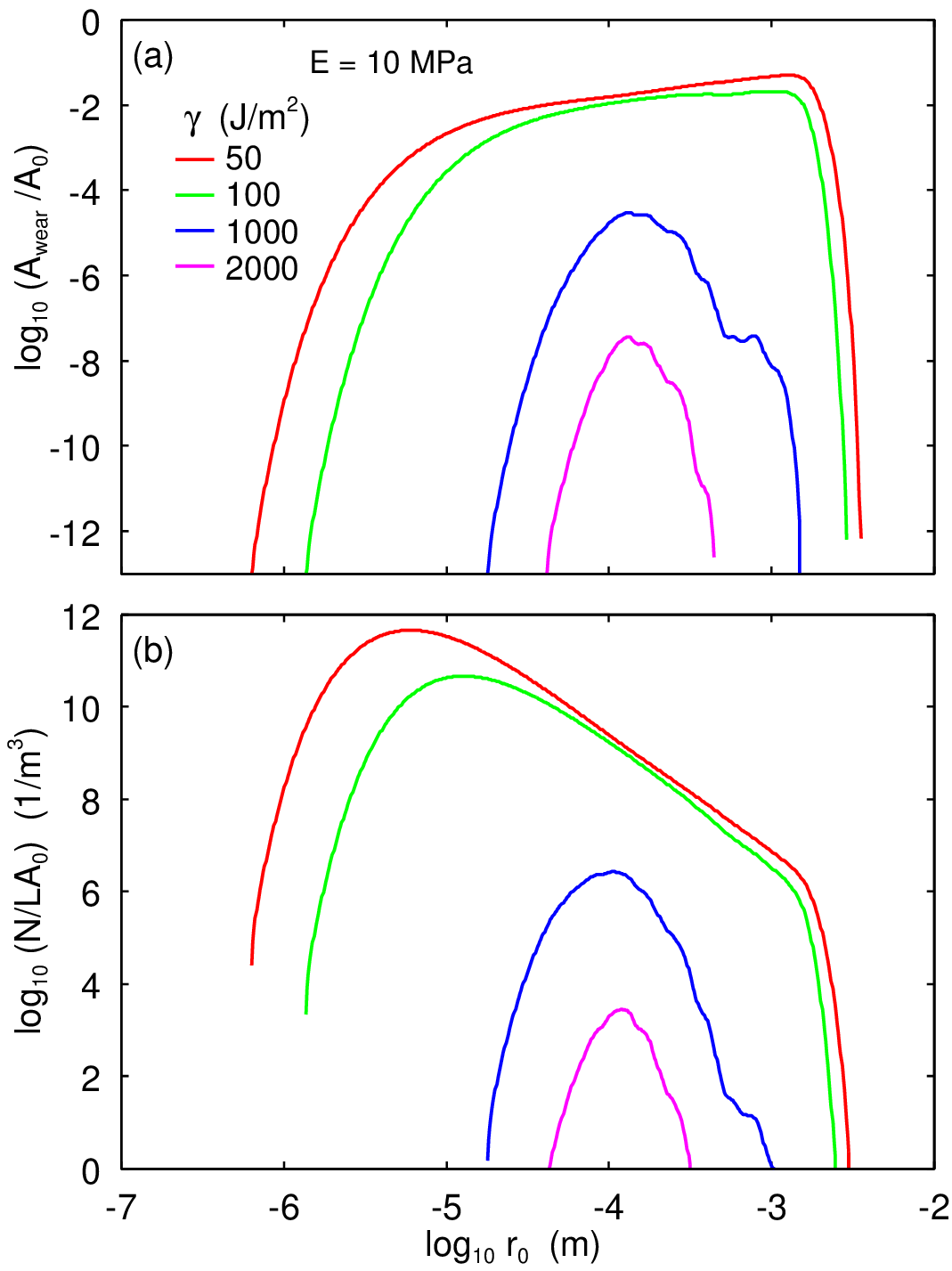}
\caption{\label{1logr0.2logNwearParticles.Many.eps}
(a) The the wear area $A_{\rm wear}$, and (b) 
the number of wear particles per unit sliding distance when $N_{\rm cont}=1$,
as a function of the radius of the wear particle (log-log scale).
In the wear area, the frictional shear stress is high enough to remove particles of size $r_0$
by cohesive crack propagation. Results are shown for the crack-energies $\gamma = 50$, $100$
$1000$ and $2000 \ {\rm J/m^2}$. For $\gamma = 3000 \ {\rm J/m^2}$ the wear area vanish.
For the Young's modulus $E=10 \ {\rm MPa}$ and Poisson ratio $\nu = 0.5$.
}
\end{figure}

\begin{figure}
\includegraphics[width=0.47\textwidth,angle=0.0]{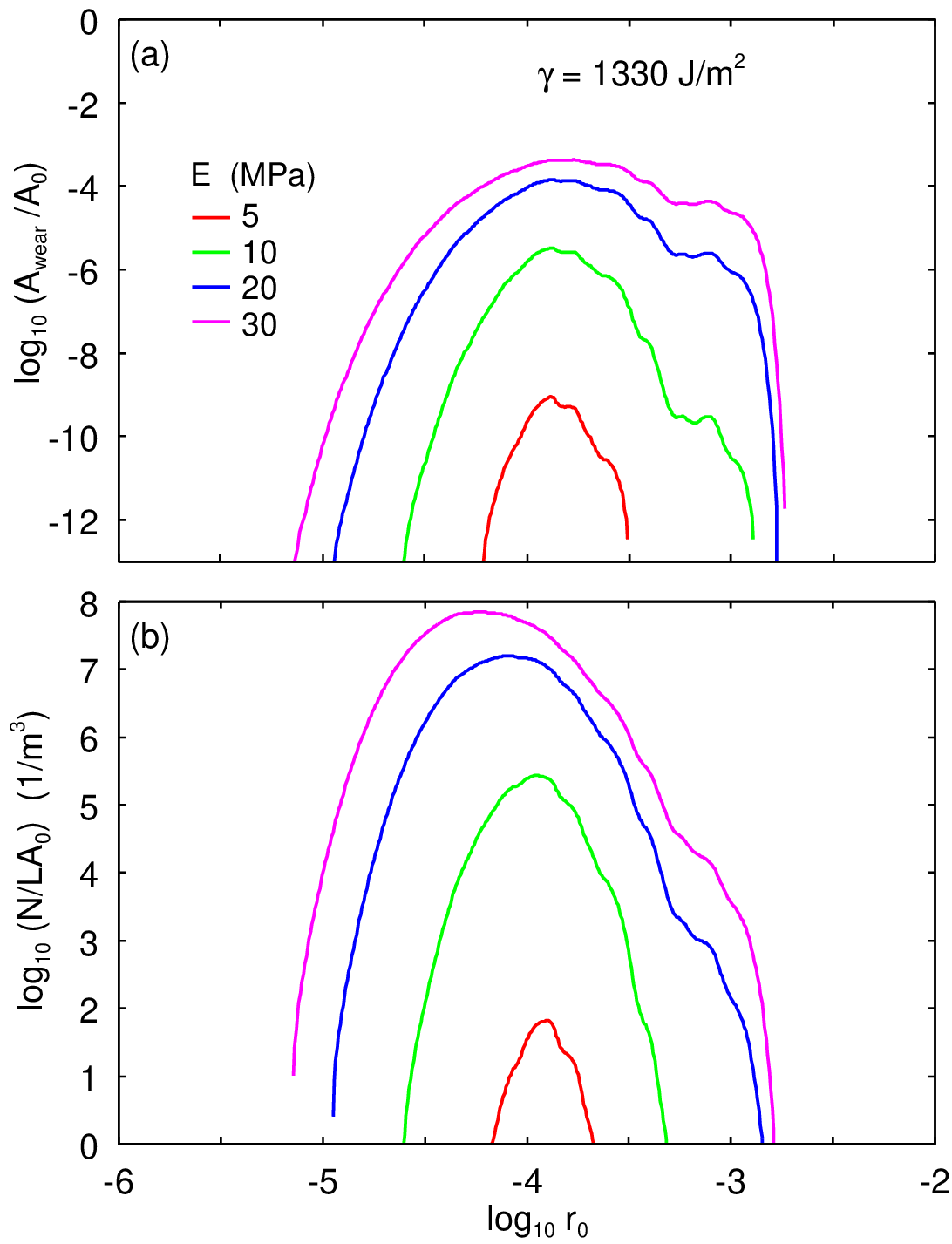}
\caption{\label{1logRadius.2logWearArea.varyE.eps}
(a) The the wear area $A_{\rm wear}$, and (b) 
the number of wear particles per unit sliding distance when $N_{\rm cont}=1$,
as a function of the radius of the wear particle (log-log scale).
In the wear area, the frictional shear stress is high enough to remove particles of size $r_0$
by cohesive crack propagation. Results are shown for the Young's modulus $E=5$, $10$, $20$ and 
$30 \ {\rm MPa}$ (with the Poisson ratio $\nu = 0.5$). The crack-energy $\gamma = 1330 \ {\rm J/m^2}$. 
}
\end{figure}

\begin{figure}
\includegraphics[width=0.47\textwidth,angle=0.0]{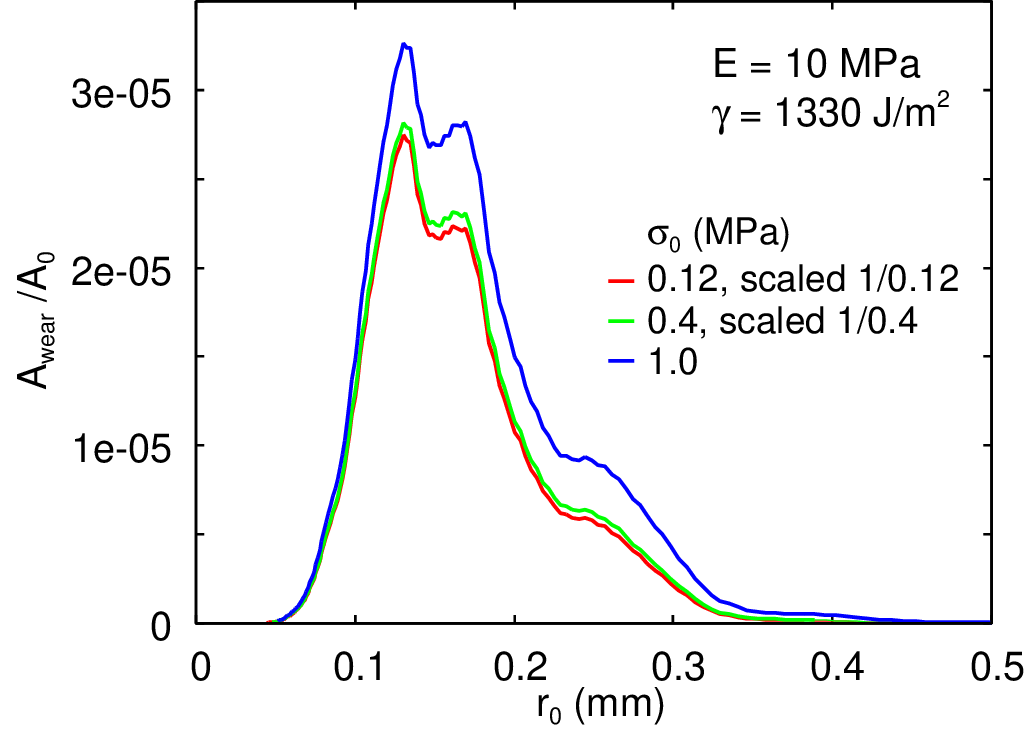}
\caption{\label{1radius.2WearAreaNormalized.eps}
The wear area $A_{\rm wear}$ in units of the nominal area $A_0$ 
as a function of the wear particle radius $r_0$ (in mm) for the nominal
contact stress $\sigma_0 =0.12$, $0.4$ and $1 \ {\rm MPa}$. We have scaled the 
wear area with $1/\sigma_0$. The figure shows that for the stress $\sigma_0 < 0.4  \ {\rm MPa}$
the wear rate is nearly proportional to $\sigma_0$ but increases slightly faster than linear for larger
contact pressures. 
}
\end{figure}

\vskip 0.3cm
{\bf 5 Numerical results}

We first present numerical results for the wear area $A_{\rm wear}$ and the number of wear particles $N$, assuming $N_{\rm cont} = 1$, as in severe wear. Unless otherwise stated, we assume the friction coefficient $\mu = 1$, Young's modulus of $E=10 \ {\rm MPa}$, crack energy of $1330 \ {\rm J/m^2}$, and nominal contact pressure of $\sigma_0 = 0.12 \ {\rm MPa}$, as in the experiments reported in Sec. 3. The assumption $N_{\rm cont} = 1$ corresponds to severe wear involving the ultimate tear strength. In the next section, where we compare the theory with the experiments, we will assume $\Delta x/r_0 \ll 1$, as expected for mild rubber wear.

Fig. \ref{1logr0.2logNwearParticles.Many.eps} shows (a) the wear area $A_{\rm wear}$ in units of the nominal area $A_0$ [Eq. (7)], and (b) the number of wear particles per unit sliding length $N/LA_0$ [Eq. (15) with $N_{\rm cont}=1$], as a function of the wear particle radius $r_0$ (log-log scale). Results are shown for crack energies $\gamma = 50$, $100$, $1000$, and $2000 \ {\rm J/m^2}$. For the number of wear particles, we assume that for each asperity contact region with radius $r_0$, one wear particle is generated during sliding over a distance equal to the diameter $2r_0$ of the contact region. Thus, $N/LA_0$ is given by (15) with $N_{\rm cont} = 1$ (or $r_0/ \Delta x = 0$).

Fig. \ref{1logRadius.2logWearArea.varyE.eps} presents the same data as in Fig. \ref{1logr0.2logNwearParticles.Many.eps} for Young's modulus of $E=10$, $20$, and $30 \ {\rm MPa}$ (with a Poisson ratio of $\nu = 0.5$) and with $\gamma = 1330 \ {\rm J/m^2}$. Note the strong dependency of the wear rate on both the crack energy $\gamma$ (Fig. \ref{1logr0.2logNwearParticles.Many.eps}) and the elastic modulus $E$ (Fig. \ref{1logRadius.2logWearArea.varyE.eps}). This dependency becomes weaker as $A_{\rm wear}$ increases, but for small $A_{\rm wear}$, it is determined by the large stress tail of the probability distribution $P(\sigma,\zeta)$, which is highly sensitive to $\sigma_{\rm c} (\zeta)$.

Fig. \ref{1radius.2WearAreaNormalized.eps} shows the wear area $A_{\rm wear}$ in units of the nominal area $A_0$ as a function of the wear particle radius $r_0$ (in mm) for nominal contact stresses of $\sigma_0 = 0.12$, $0.4$, and $1 \ {\rm MPa}$, with the wear area scaled by $1/\sigma_0$. The figure shows that for $\sigma_0 < 0.4 \ {\rm MPa}$, the wear rate is nearly proportional to $\sigma_0$, but it increases slightly faster than linear at higher contact pressures. In tire applications, the nominal pressure in the tire-road footprint is typically below $0.5 \ {\rm MPa}$, so we expect a wear rate that is roughly linear with the load for low slip velocities (e.g., $v < 1 \ {\rm mm/s}$), where frictional heating is negligible.

\vskip 0.3cm
{\bf 6 Comparison with experiments}

We have found that the wear rate is independent of sliding speed, which is remarkable. Although slower speeds mean longer contact times between the rubber and substrate asperities, this does not result in greater crack propagation distances as one might anticipate. This independence arises because cracks propagate in discrete steps rather than continuously; specifically, they follow a stick-slip pattern. During sliding, strain energy accumulates at the crack tip until it reaches a critical threshold, after which the crack advances incrementally by a displacement $\Delta x$. This process results in a wear rate governed primarily by the cumulative number of asperity contacts per unit sliding distance, rather than by contact time. Consequently, the crack tip moves at a speed unrelated to the relative speed of rubber-asperity contact, leading to a wear rate that remains constant across the tested range of sliding speeds.

Another case where one would expect a wear rate independent of sliding speed is if wear particles form at each contact where the stored elastic energy exceeds that required to form a particle. In this case, the wear rate would depend only on the sliding distance and not on the sliding time. Formation of a wear particle in each contact with a wear-asperity corresponds to severe wear and may occur on surfaces with very sharp roughness, where the elastic energy release rate matches the ultimate tear strength $\gamma_{\rm c}$ (see Fig. \ref{TearStrength.eps}). However, this does not occur on the concrete surface used in this study.

The lowest crack energy $\gamma$ for tire tread rubber is typically $\sim 10^2 \ {\rm J/m^2}$, and the highest $\sim 10^4-10^5 \ {\rm J/m^2}$. Fig. \ref{1logr0.2logNwearParticles.Many.eps} shows that rubber wear particles are expected to range in size from ${\rm \mu m}$ to ${\rm mm}$, which agrees with experimental observations. Tire wear particles have also been observed in the ${\rm nm}$-size range, but these particles cannot result from unperturbed crack propagation and must result from a different physical process than considered here, such as slow crack motion due to reactions with foreign molecules (e.g., oxygen or ozone) at the crack tip (stress corrosion).

We will now compare the theoretical predictions to the experimental results for the bus-truck tread compound, which is based on natural rubber. For this compound, we measured a wear rate of $\approx 2.5 \ {\rm mg/m}$ (see Fig. \ref{1l.3dm.for.all.velocities.eps}). Assuming a rubber mass density of $1250 \ {\rm kg/m^3}$, this gives a wear volume per unit sliding distance of $V/L \approx 2 \ {\rm mm^3/m}$. In the experiment, the nominal contact pressure was $\sigma_0 \approx 0.12 \ {\rm MPa}$, the normal force was $250 \ {\rm N}$, and the nominal contact area was $A_0 = F_{\rm N}/\sigma_0 \approx 0.002 \ {\rm m^2}$. In the calculations below for the concrete surface, we use friction coefficient $\mu=0.9$  
unless otherwise stated. This is a typical friction coefficient observed on concrete, see, e.g., Fig. \ref{1slidingDistance.2mu.all.loads.eps}.

To compare the measured wear rate to the theory prediction, we need to know the effective modulus $E$, which depends on the sliding speed (see Appendix B). The characteristic deformation frequency when a rubber block slides in contact with a road asperity is $\omega \approx v/r_0$, where $\sim r_0$ is the linear size of the contact region. In the present case, the contact radius $\approx 0.1 \ {\rm mm}$, and the sliding speeds range from $v=1-10^4 \ {\rm \mu m/s}$, giving deformation frequencies between $0.01 \ {\rm s}^{-1}$ and $100 \ {\rm s}^{-1}$. In this frequency range, the low-strain modulus varies from approximately $27$ to $36 \ {\rm MPa}$. For the large strains relevant here, 
we show in Appendix B that the effective (secant) modulus is $E \approx 15 \pm 5 \ {\rm MPa}$.

\begin{figure}
\includegraphics[width=0.47\textwidth,angle=0.0]{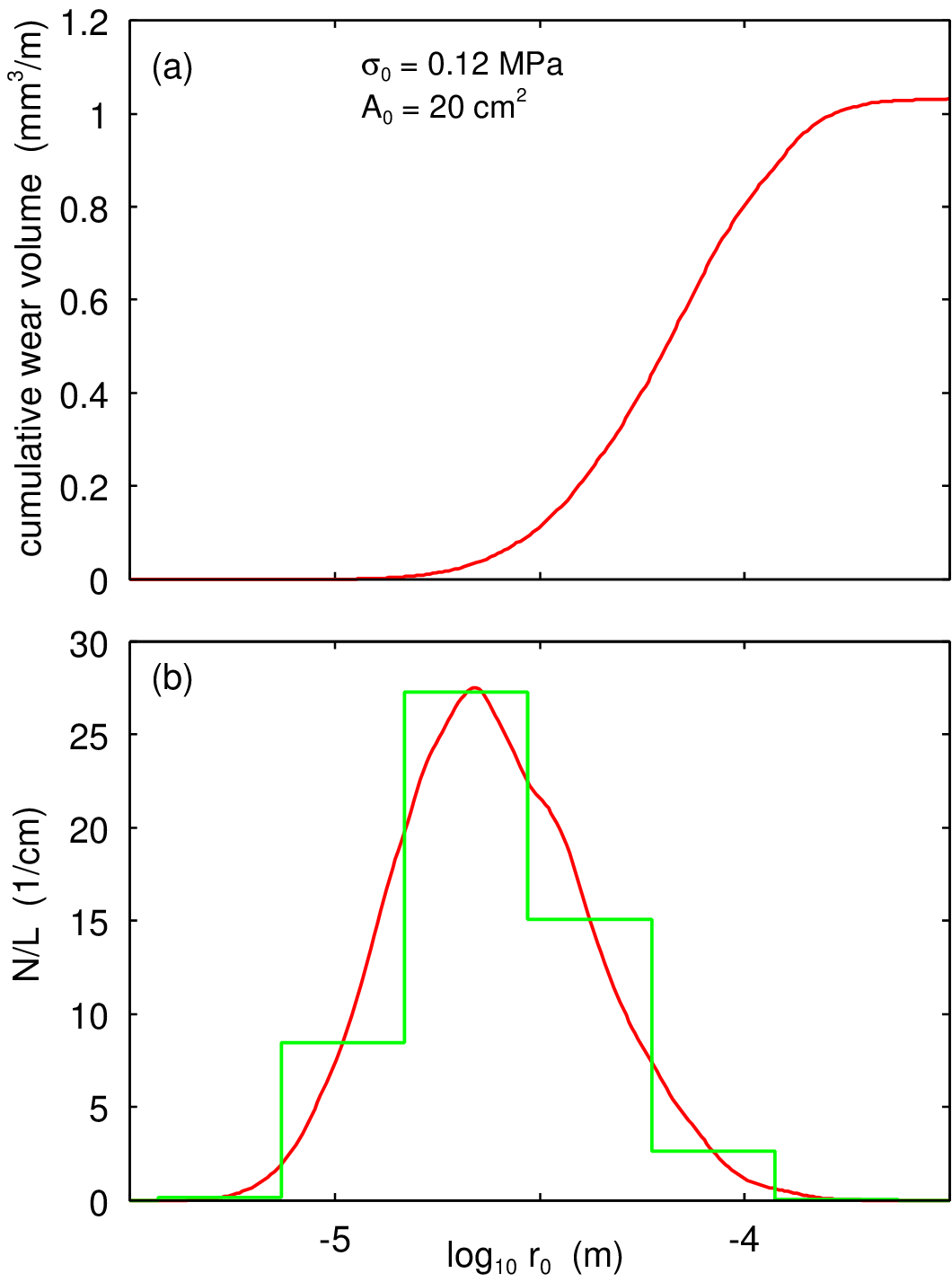}
\caption{\label{1logr0.2Number.concrete.new.eps}
The cumulative wear volume (a), and the number of generated particles (b), 
as a function of the logarithm of the particle radius,
for the same system as studied experimentally where the wear volume is $\approx 2 \ {\rm mm^3/m}$.
We have used $E=10 \ {\rm MPa}$, $\nu = 0.5$ and the measured relation between the crack-tip displacement $\Delta x (\gamma)$ 
and the tearing energy $\gamma$ shown in Fig. \ref{1logGamma.2WearIntegrand.redConcrete.GreenP100.eps}(a).
The nominal contact area $A_0 = 0.002 \ {\rm m^{-2}}$ and the nominal contact pressure $\sigma_0 = 0.12 \ {\rm MPa}$
as in the experiment in Sec. 3.
}
\end{figure}

\begin{figure}
\includegraphics[width=0.47\textwidth,angle=0.0]{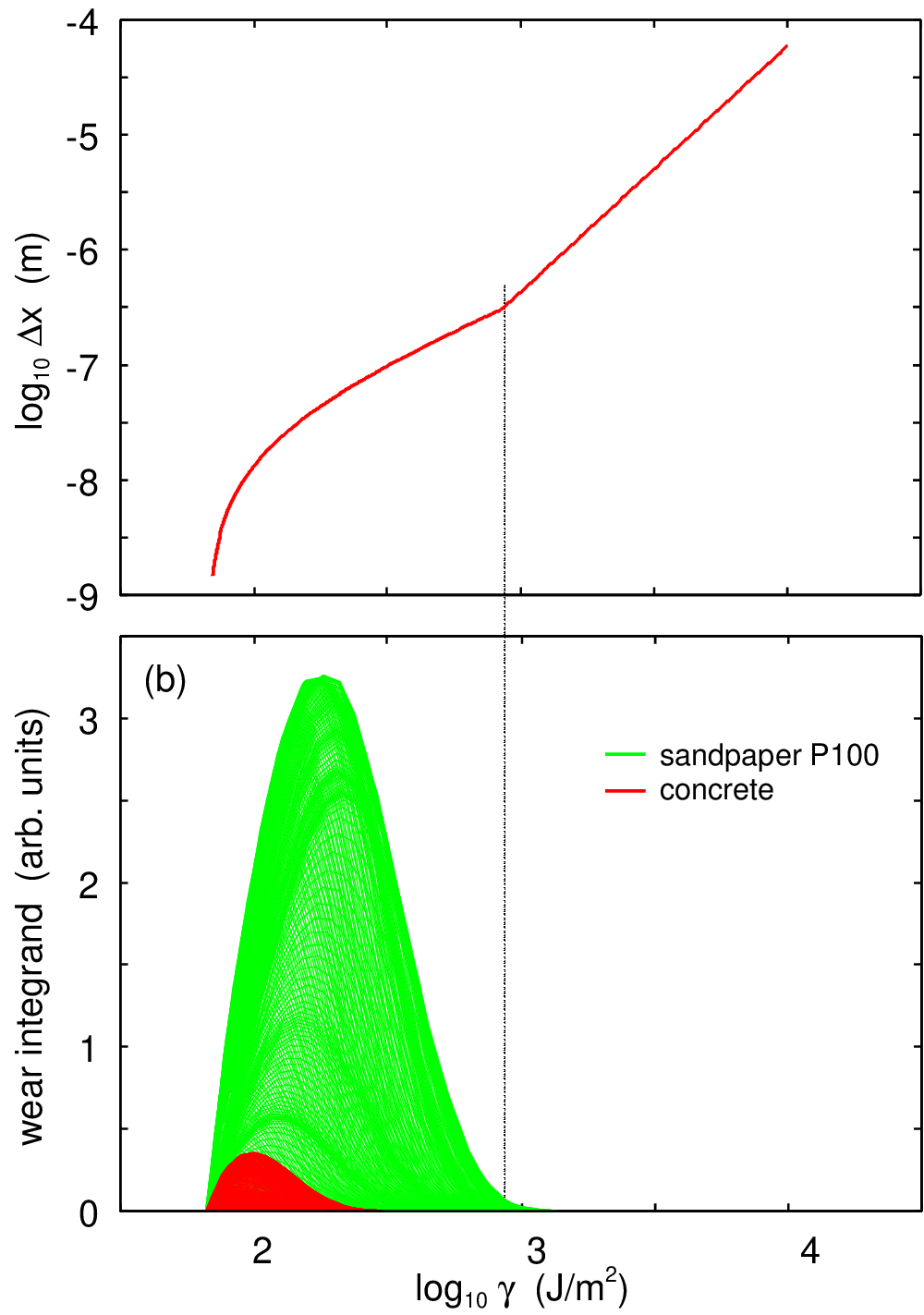}
\caption{\label{1logGamma.2WearIntegrand.redConcrete.GreenP100.eps}
(a) The relation between the crack tip displacement per oscillation and the tearing (or crack) energy $\gamma$.
(b) The integrand in (20) as a function of $\gamma$ for all magnifications
(or particle radius $r_0$) for concrete (red) and for the sandpaper P100 (green). 
[Note: the integration variable in (20) is the pressure but for each pressure corresponds to
a tearing energy as given by (13).] The red and green area is the superposition of many curves for the different 
magnifications or particle radius.
We have used $E=10 \ {\rm MPa}$, $\nu = 0.5$ and the relation between the crack-tip displacement $\Delta x (\gamma)$ 
and the tearing energy $\gamma$ shown in (a).
The nominal contact area $A_0 = 0.002 \ {\rm m^{-2}}$ and the nominal contact pressure $\sigma_0 = 0.12 \ {\rm MPa}$
as in the experiment in Sec. 3. We have used the friction coefficient $\mu = 0.9$ for concrete and $1.06$ for sandpaper.
}
\end{figure}

Using the measured power spectrum (see Appendix C) and the elastic modulus $E=10 \ {\rm MPa}$ (with $\nu = 0.5$), Fig. \ref{1logr0.2Number.concrete.new.eps} shows (a) the cumulative wear volume and (b) the number of generated particles as a function of the logarithm of the particle radius. Here, we have used the general 
relationship between the crack-tip displacement $\Delta x$ and the tearing energy $\gamma$ observed in experiments, 
which is well approximated by

$$\Delta x = 0 \ \ \ \ \ \ {\rm for} \ \gamma < \gamma_0$$
$$\Delta x = a\times (\gamma-\gamma_0), \ \ \ \ \ \ {\rm for} \ \gamma_0 < \gamma < \gamma_1$$
$$\Delta x = b\times T^\alpha  \ \ \ \ \ \ {\rm for} \ \gamma_1 <\gamma < \gamma_2 .$$

Here, $\gamma_1$ is defined as the tearing energy where the last two expressions for $\Delta x$ are equal, and as $\gamma$ increases above $\gamma_2 \approx 10^5 \ {\rm J/m^2}$, then $\Delta x$ rapidly approaches $\infty$. 
We use (in SI units), $a=3.9 \times 10^{-10}$, 
$\gamma_0 = 66.24$, $b=1.6\times 10^{-13}$, and $\alpha = 2.14$. 

The particle size distribution shown in Fig. \ref{1logr0.2Number.concrete.new.eps}(b) is consistent with the optical image of the wear particles in Fig. \ref{RubberParticle2.ps}. Thus, the simple theory presented above aligns with the experimental observations in terms of both the wear rate ($\sim 2 \ {\rm mm^3/m}$) and the observed particle sizes ($\sim 3-100 \ {\rm \mu m}$).

The small deviation between the calculated and measured wear rate may be due to approximations in the theory, e.g., the factor $\beta$ in (4) is not exactly 1 as assumed above, and the way we separate particle sizes in steps of factors of 2 is non-unique. Additionally, the relationship between $\gamma$ and $\Delta x$ for micrometer-sized cracks may differ from that of macroscopic cracks, e.g., the frequency of pulsating deformations will differ (see Sec. 7).

In Fig. \ref{1logGamma.2WearIntegrand.redConcrete.GreenP100.eps}(a), we show the used relationship between the crack tip displacement 
$\Delta x$ per oscillation, and the tearing (or crack) energy $\gamma$ (log-log scale). In Fig. \ref{1logGamma.2WearIntegrand.redConcrete.GreenP100.eps}(b), we show the integrand in (21) as a function of $\gamma$. The figure consists of numerous curves for different magnifications (or wavenumber cut-offs), corresponding to different particle sizes.

The integration variable in (21) is the pressure, but each pressure corresponds to the tearing energy as given by (13). The red area is the superposition of many curves for different magnifications or particle radii. 
Also shown is the result for the sandpaper P100 to be studied in the next section. The sandpaper gives much higher wear rates
and involves larger tearing energies or, equivalently, larger $\Delta x$ corresponding to faster crack propagation. Note that both
cases only involves the tearing energy in the region where $\Delta x$ increases linearly with  $\gamma$.  

For the concrete surface
the center-of-mass of the curves is around $\gamma \approx 100 \ {\rm J/m^2}$, relatively close to the fatigue threshold tearing strength ($\gamma_0 \approx 66 \ {\rm J/m^2}$), as expected for mild rubber wear. For surfaces with sharper roughness, such as sandpaper, 
the elastic energy stored in the asperity contact regions is greater, and particle removal involve higher 
tearing energies and much higher wear rates.

\begin{figure}
\includegraphics[width=0.47\textwidth,angle=0.0]{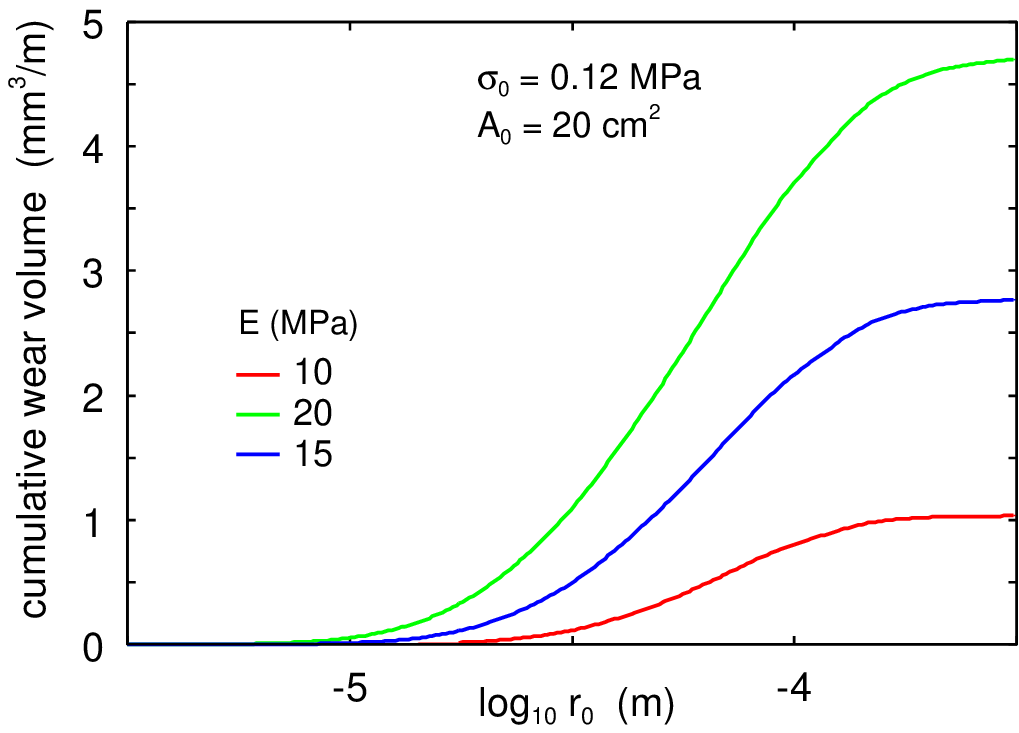}
\caption{\label{1logr0.2WearVolume.dependOnE.eps}
The cumulative wear volume as a function of the logarithm of the particle radius,
for the same system as in Fig. \ref{1logr0.2Number.concrete.new.eps} for $E=10 \ {\rm MPa}$ (red curve),
$E=20 \ {\rm MPa}$ (green) and $E=15 \ {\rm MPa}$ (blue).
The other parameters are the same as in Fig. \ref{1logr0.2Number.concrete.new.eps}.
}
\end{figure}

We will now show how the wear rate depends on the different parameters that enter in the theory.
We note that it is in general not possible to design experiments where only one material 
property is changed, e.g., it is not possible
to modify the elastic modulus without changing the relation between the tearing energy $\gamma$ and $\Delta x$ so the
theory results presented here may not be easy to test experimentally.

\begin{figure}
\includegraphics[width=0.47\textwidth,angle=0.0]{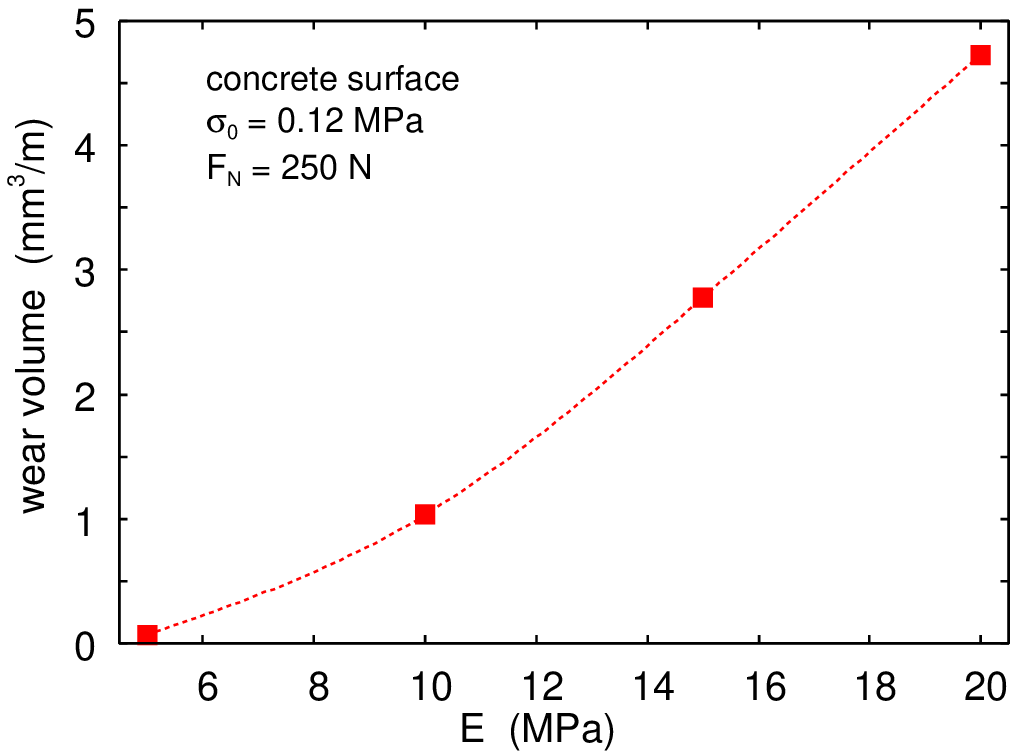}
\caption{\label{1E.2wearrate.for.concrete.eps}
The calculated wear rate as a function of the elastic modulus $E$ (with $\nu = 0.5$)
for rubber block sliding on the concrete surface.
The wear rate for $E= 5 \ {\rm MPa}$ is $0.068 \ {\rm mm^3/m}$. 
}
\end{figure}

Fig. \ref{1logr0.2WearVolume.dependOnE.eps} shows the same results as in Fig. \ref{1logr0.2Number.concrete.new.eps}(a), for the same parameters except for the green and blue curve, which are for $E=20$ and $15 \ {\rm MPa}$. 
In all the calculations we used the $\Delta x (\gamma)$ relation shown in Fig. \ref{1logGamma.2WearIntegrand.redConcrete.GreenP100.eps}(a).
For $E=15 \ {\rm MPa}$ the calculated wear rate is very close to the measured value. 
Fig. \ref{1E.2wearrate.for.concrete.eps} shows the
calculated wear rate as a function of the elastic modulus $E$ (with $\nu = 0.5$)
for the rubber block sliding on the concrete surface.

\begin{figure}
\includegraphics[width=0.47\textwidth,angle=0.0]{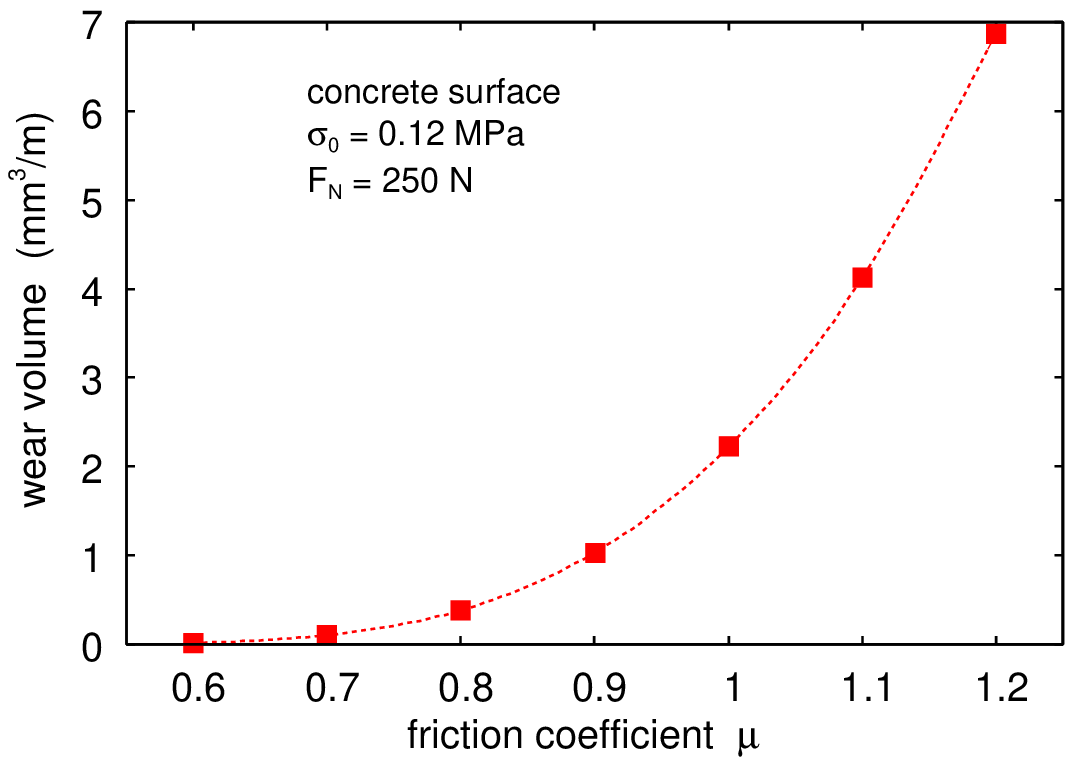}
\caption{\label{1mu.2wearrate.Concrete.eps}
The calculated wear rate as a function of the friction coefficient for rubber block sliding on a surface with
the same power spectrum as the concrete surface. The wear rate for $\mu = 0.6$ and $0.7$ are
$0.016$ and $0.103 \ {\rm mm^3/m}$, respectively. 
}
\end{figure}

\begin{figure}
\includegraphics[width=0.47\textwidth,angle=0.0]{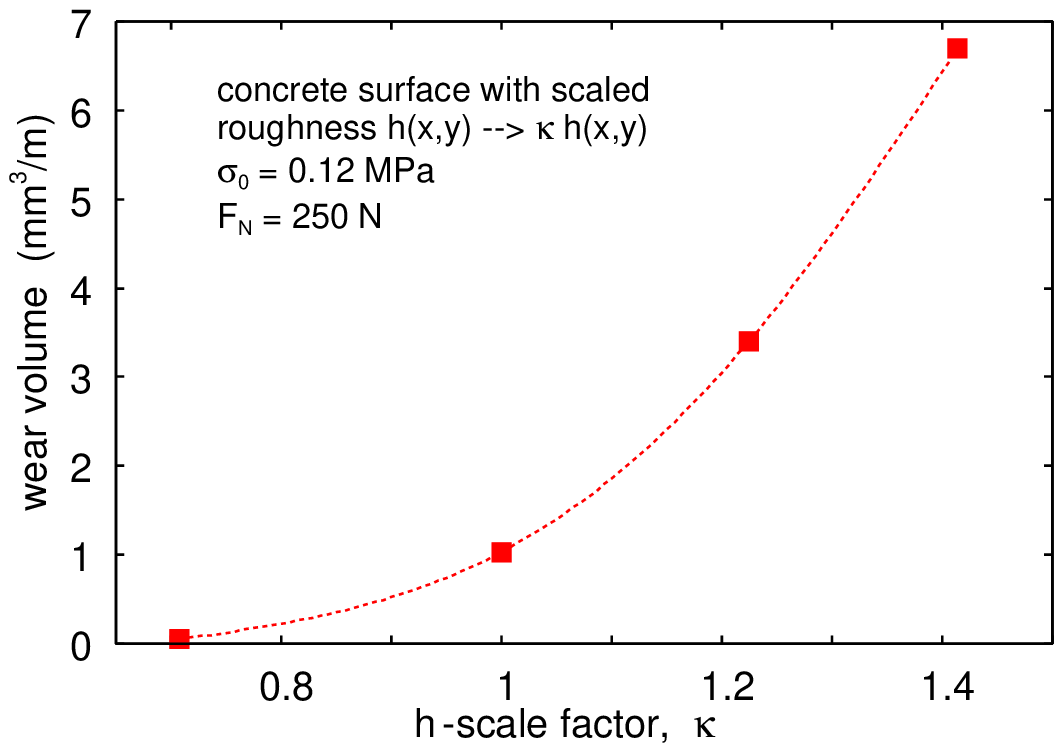}
\caption{\label{1scaleRMS.2WearRate.usingCqConcrete.eps}
The calculated wear rate as a function of the surface roughness amplitude. The surface roughness of the
concrete surface is scaled by a factor $\kappa$, $h(x,y) \rightarrow \kappa h(x,y)$ 
Thus we have used in the calculations the $C(q)$ of the concrete surface scaled
by $\kappa^2$. The wear rate for $\kappa = 0.707$ is $0.048 \ {\rm mm^3/m}$.
}
\end{figure}

Fig. \ref{1mu.2wearrate.Concrete.eps} and  \ref{1scaleRMS.2WearRate.usingCqConcrete.eps}
shows the dependency of the wear rate on the friction coefficient, and on the magnitude of the surface roughness.
In the latter case, we have scaled the height profile with the indicated number $\kappa$, which corresponds to
scaling the power spectrum of the concrete surface with the factor $\kappa^2$.
Note the strong variation of the wear rate with the parameters $E$, $\mu$, and $\kappa$ in
Fig. \ref{1E.2wearrate.for.concrete.eps} - \ref{1scaleRMS.2WearRate.usingCqConcrete.eps}.

\begin{figure}
\includegraphics[width=0.47\textwidth,angle=0.0]{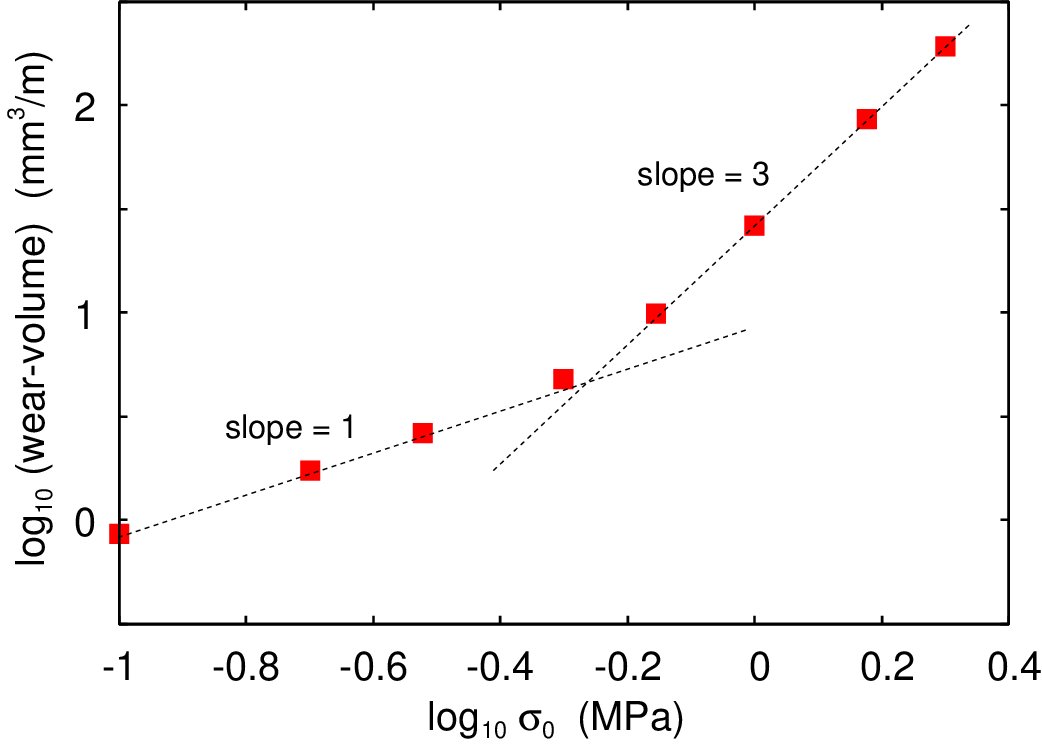}
\caption{\label{1logSigma0.2logWearRate.eps}
The calculated wear rate as a function of the nominal contact pressure (log-log scale) 
for rubber block sliding on the concrete surface. 
}
\end{figure}

\begin{figure}
\includegraphics[width=0.47\textwidth,angle=0.0]{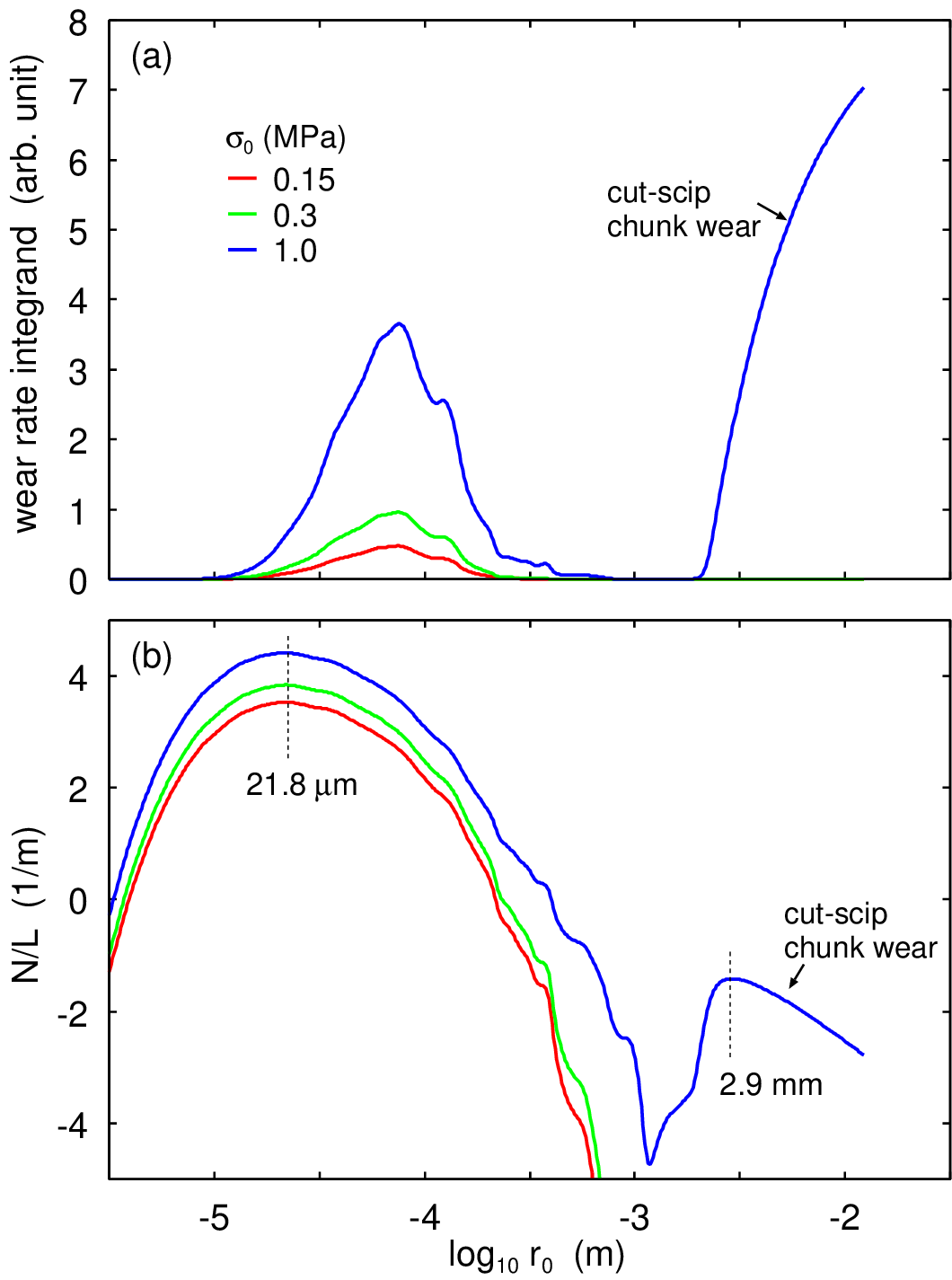}
\caption{\label{1logr0.2WearRateIntegrand.eps}
(a) The integrand in the $\xi$-integral in the wear volume integral (21) as a function of the
logarithm of the radius $r_0$ of the wear particle. 
(b) The particle distribution assuming the nominal contact area $A_0 = 20 \ {\rm cm}^2$.
The sliding distance to remove a particle in a 2-interval around $r_0 \approx 2.9 \ {\rm mm}$
is about $26 \ {\rm m}$ and to remove a particle in a 2-interval around $r_0 \approx 1 \ {\rm cm}$
the sliding distance is nearly $1 \ {\rm km}$.
The results are for the concrete surface for 
the nominal contact pressures $\sigma_0 = 0.15$, $0.30$ and $1.0 \ {\rm Pa}$.
}
\end{figure}

\begin{figure}
\includegraphics[width=0.47\textwidth,angle=0.0]{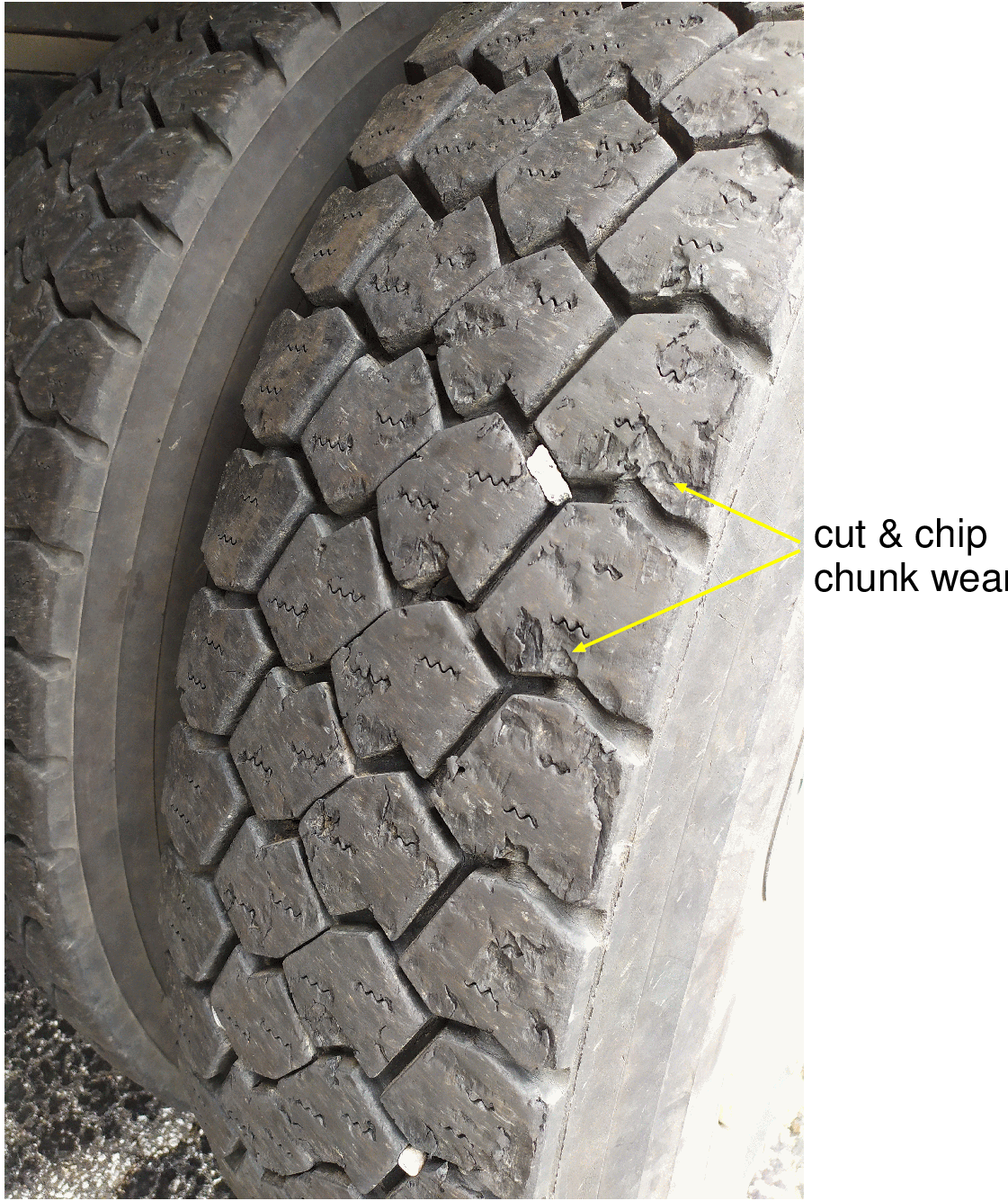}
\caption{\label{cutANDchip.eps}
Cut \& chip wear of truck tire result from removing macroscopic ($\sim {\rm cm}$) patches of
rubber from the tire surface when driving on very rough road surfaces with long wavelength
or large scale ($\sim {\rm cm}$) inhomogeneities such as gravel or roots
(R. Stocek, private communication).
}
\end{figure}

Fig. \ref{1logSigma0.2logWearRate.eps} 
shows the calculated wear rate as a function of the nominal contact pressure (log-log scale) for the rubber block sliding on the concrete surface. For contact pressures $\sigma_0 < 0.4 \ {\rm MPa}$, the wear rate is proportional to the contact pressure and, therefore, to the normal (loading) force. 
For higher pressures, however, the wear rate increases rapidly as $\sigma_0^3$. 
This is primarily due to the formation of very large wear particles, in addition to the $1-100 \ {\rm \mu m}$ particles formed at lower nominal contact pressures. 
This additional contribution to the wear rate is referred to as cut-chip-chunk (CCC) wear\cite{CCC1,CCC2,CCC3,CCC4}. 
It occurs because elastic energy scales with the size of the system as $r_0^3$, while the fracture energy is proportional to $r_0^2$; thus, at large enough length scales, there will always be more elastic energy stored than needed to form wear particles.

To illustrate that the additional contribution is due to the formation of large wear particles, in Fig. \ref{1logr0.2WearRateIntegrand.eps}(a) we show the integrand in the $\xi$-integral in the wear volume integral (21), and in (b) the number distribution of wear particles, 
as a function of the logarithm of the radius $r_0$ of the wear particle. 
Results are shown for the concrete surface at nominal contact pressures $\sigma_0 = 0.15$, $0.30$, and $1.0 \ {\rm MPa}$. 

The CCC contribution to the wear volume
increases with the wear particle radius up to the cut-off determined by the smallest wavenumber $q_0$ for which the surface roughness power spectrum was determined, $(r_0)_{\rm max} = \pi/q_0$. Since $q_0$ is defined by the scan length $L$ of the topography measurement, $q_0 = \pi /L$, it follows that $(r_0)_{\rm max} = L$. In reality, there is always some physical length scale that determines the largest rubber fragments removed. For tires, this may be the size of the tread blocks or the thickness of the rubber layer on the steel cord of slick tires (see below).

Since CCC wear involves removing macroscopic chunks of rubber, the relationship between the crack tip displacement $\Delta x$ and the tearing energy $\gamma$ measured for macroscopic rubber samples may be more relevant than the relationship used in this study for the removal of small ($\sim 10 \ {\rm \mu m}$) particles, where the displacement $\Delta x$ is assumed to be enhanced by a reduction in viscoelastic screening and strain crystallization.

It is also worth noting that tires are designed such that the nominal pressure when driving on normal road surfaces is not high enough for CCC wear to occur. However, when driving off-road, inhomogeneities such as gravel or roots can generate nominal pressures high enough to induce CCC wear. This is often the case for truck tires in off-road applications (see Fig. \ref{cutANDchip.eps}), and it is also commonly observed in conveyor belts.

Studying the wear integrand as a function of $\gamma$, as in Fig. \ref{1logGamma.2WearIntegrand.redConcrete.GreenP100.eps}(b), shows that the 
CCC wear rate involves roughly the same range of $\gamma$ values ($\gamma < 300 \ {\rm J/m^2}$) for both $\sigma_0 = 0.30$ and $1.0 \ {\rm MPa}$. This indicates that the CCC wear fragments involve similar crack tip displacements $\Delta x$ as those required to remove much 
smaller wear particles. Hence large wear fragments will be removed very infrequently, but may still correspond 
to the largest wear mass. 

In Fig. \ref{1logr0.2WearRateIntegrand.eps}(b), we show the number of wear particles produced, assuming a nominal contact area of $A_0 = 20 \ {\rm cm}^2$. Note that for all nominal contact pressures, the $N/L$ curves have maxima at the same particle radius, $r_0 \approx 21.8 \ {\rm \mu m}$. In fact, the wear particle distribution in the $1-100 \ {\rm \mu m}$ size range appears not to depend on the contact pressure, except for a scaling with the magnitude of the applied normal force. This observation is in qualitative agreement with optical images of particle distributions at different normal loads.

Fig. \ref{1logr0.2WearRateIntegrand.eps}(b) also shows that the sliding distance required to remove a particle in a 2-interval around $r_0 \approx 2.9 \ {\rm mm}$ is about $26 \ {\rm m}$, while removing a particle in a 2-interval around $r_0 \approx 1 \ {\rm cm}$ requires nearly $1 \ {\rm km}$ of sliding. In the calculations, we used the relationship between $\gamma$ and $\Delta x$ shown in Fig. \ref{1logGamma.2WearIntegrand.redConcrete.GreenP100.eps}.
However, the CCC wear region involves 
macroscopic sized cracks, where $\Delta x$ is smaller, so even larger sliding distances than those calculated above may be needed to remove large chunks of rubber.

To summarize, there are three size regions in rubber wear:

- At very short length scales (nanoscale), the stored elastic energy in asperity contact regions is insufficient to propagate cracks and remove nanoscale wear particles. In this regime, wear particles are produced by stress corrosion. In stress corrosion, molecules from the surrounding atmosphere, adsorbed molecules, or thin contamination films react with rubber chains that are stretched due to the frictional shear stress. This stretching lowers the barriers for chemical reactions, enabling a stress-aided thermally activated process.

- At length scales of around $1-100 \ {\rm \mu m}$, the elastic deformation energy in some asperity contact regions during slip exceeds the fracture energy, i.e., the energy per unit area needed to break the bonds at the crack tip. This generates wear particles on the length scale of $1-100 \ {\rm \mu m}$.

- Since the stress required to propagate cracks scales as $1/r_0$ with the size of the stressed region, see (3) and (4), it follows that at large enough length scales (typically $\sim 1 \ {\rm cm}$), $\sigma_{\rm c}$ is below the nominal shear stress, allowing the wear of removing large chunks of rubber. That is, at large enough length scales, the elastic deformation energy $\sim r_0^3$ is larger than the fracture energy $\sim r_0^2$, enabling crack propagation to remove a rubber fragment. The scaling of the stored elastic energy $\sim r_0^3$ only holds for systems with linear sizes larger than $r_0$, and in real applications, some physical constraints will determine (or limit) the size of the removed rubber fragments.

The removal of large chunks of rubber, i.e., cut-chip-and-chunk (CCC) wear, is well-known for tires and conveyor belts and is included in our rubber wear theory. In an infinite system, the removed rubber fragments could theoretically be infinitely large, but in real applications, some physical constraints limit their size. For tires, for example, one would expect the fragments to be smaller than the tread block size or, in the case of slick tires, the thickness of the rubber layer covering the steel cord.

\begin{figure}
\includegraphics[width=0.47\textwidth,angle=0.0]{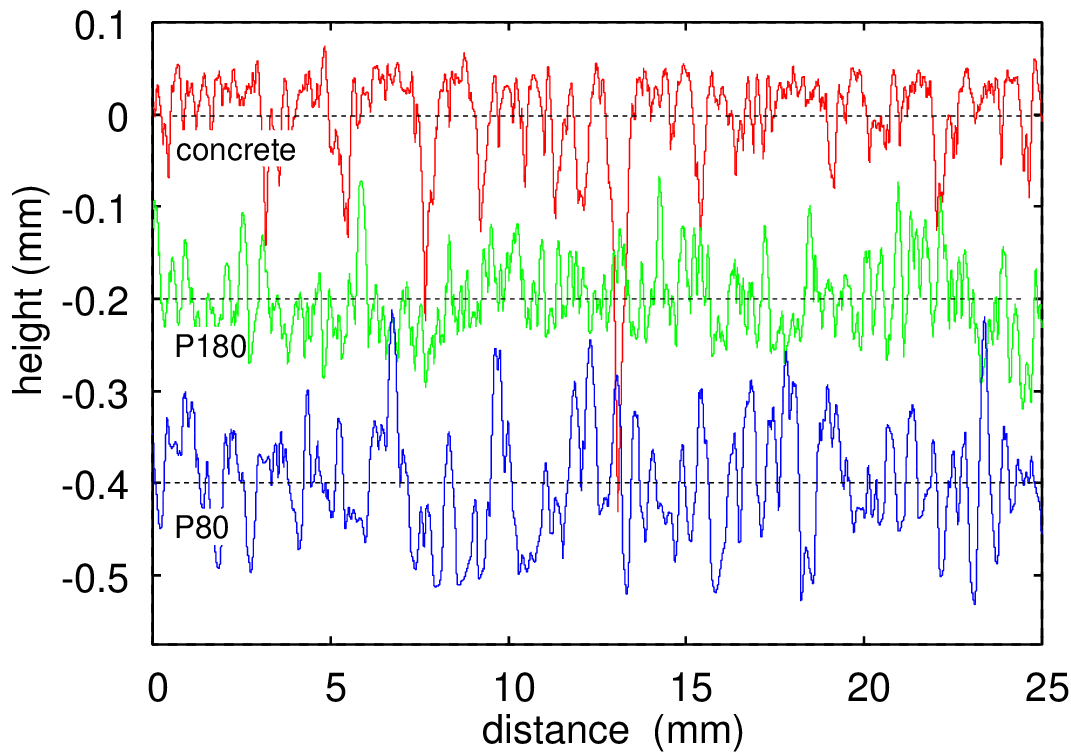}
\caption{\label{1x.2h.concrete.sandpaper180.80.eps}
The measured surface height $h(x)$ as a function of the distance $x$ along a strait
25 mm long track on the concrete surface (red) and on the sandpaper
P80 (green) and P180 (blue) surfaces. Note that the upper part of the height
profile on the concrete surface is smoother than the lower part.
The horizontal dashed lines are the average surface plane $\langle h \rangle = 0$
and the data for the P80 and P180 surfaces are shifted downwards by $0.2 \ {\rm mm}$
and $0.4 \ {\rm mm}$, respectively.
}
\end{figure}

\begin{figure}
\includegraphics[width=0.47\textwidth,angle=0.0]{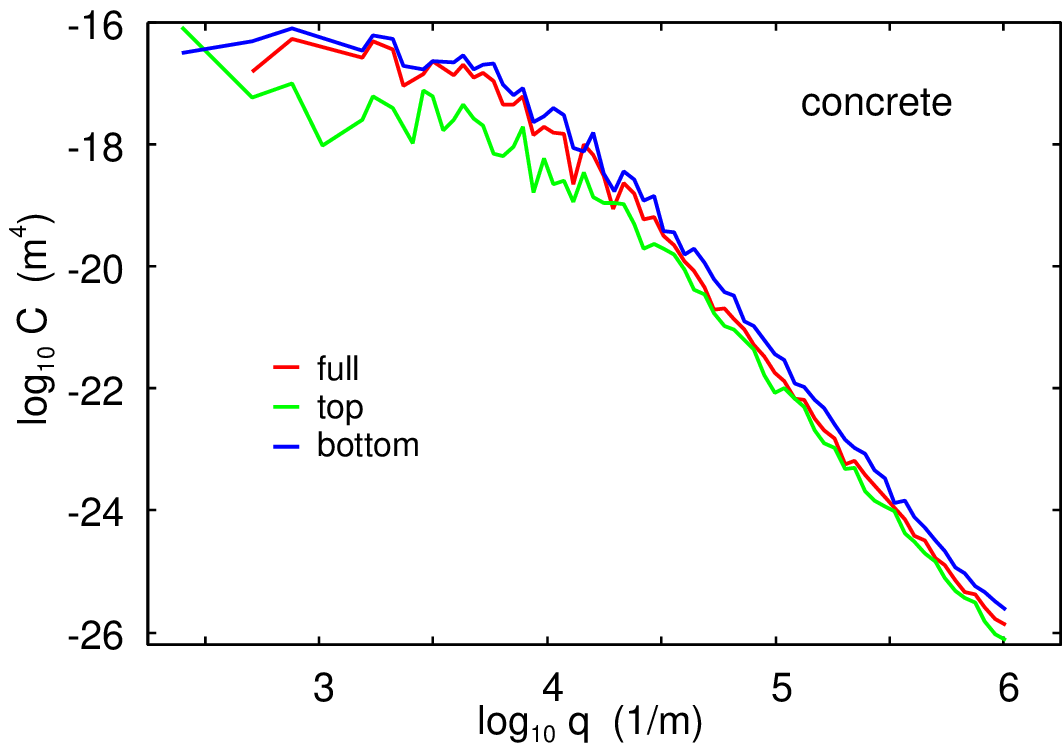}
\caption{\label{1logq.2logC.concrete.T.B.Full.eps}
The full (red), bottom (blue), and top (green) surface roughness power 
spectrum as a function of the wavenumber for
the concrete surface.
}
\end{figure}

\begin{figure}
\includegraphics[width=0.47\textwidth,angle=0.0]{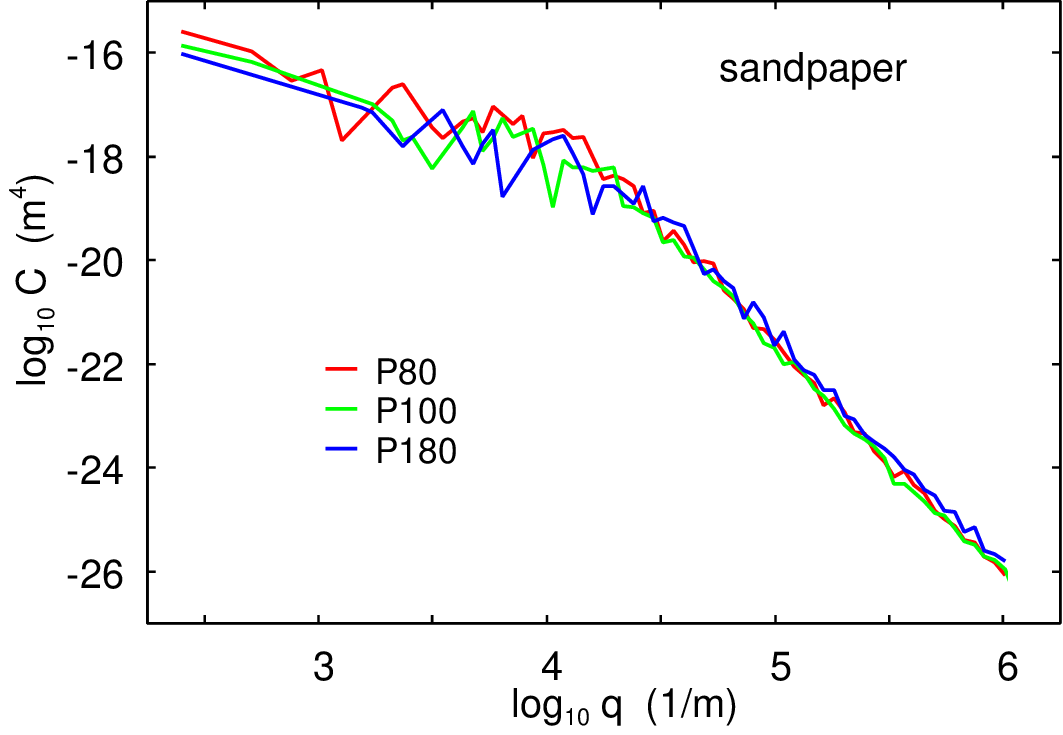}
\caption{\label{1logq.2logC.P80.P100.P180.eps}
The surface roughness power
spectrum as a function of the wavenumber for the sandpaper P80 (red), P100 (green), and P180 (blue)
surfaces.
}
\end{figure}

\vskip 0.3cm
{\bf 7 Additional test of the theory}

After the study reported above was completed, we decided to further test the theory by measuring the wear rate on several different sandpaper surfaces. It is known that rubber wear rates on sandpaper are much higher than on road surfaces, and a crucial test of the theory is to determine if it can account for this difference.

Sandpaper has sharper roughness than the concrete surface we used, which is easily observed by sliding a finger over the concrete and sandpaper surfaces. But why is this the case despite the fact that the rms-roughness of the concrete surface is similar to that of the sandpaper surfaces ($59 \ {\rm \mu m}$ for the concrete surface, and $32$, $31$, and $47 \ {\rm \mu m}$ for P180, P100, and P80 sandpaper surfaces, respectively)?. Both concrete and sandpaper consist of particles bound together with a binder (epoxy resin for sandpaper and calcium silicate hydrates for concrete), but the concrete surface feels smoother to the touch than sandpaper because it is molded against a flat surface, resulting in a surface where the tops of the stone particles are at (nearly) equal heights, while this is not the case for sandpaper. In sandpaper, particles are deposited (electrostatically) as a monolayer on top of the nominally flat surface covered by the resin binder. The particles have various shapes and sizes (with sandpaper particle diameters typically fluctuating by approximately $50\%$ around their average value), resulting in a surface where the particle heights fluctuate by a similar amount as the average particle size.

Newly prepared asphalt road surfaces (which consist of stone particles with a bitumen binder) are similarly smooth to the concrete surface used in this study. This results from the use of heavy rollers, which deform the asphalt surface so that the tops of the stone particles at the top surface are at (nearly) the same height.

We will refer to surfaces, such as the concrete surface, where the tops of the highest asperities are of (nearly) equal height as having {\it smooth roughness}, while surfaces, where the tops fluctuate randomly, will be described as having {\it sharp roughness}. Note that smooth or sharp roughness is unrelated to the rms-roughness; for example, the rms-roughness of the concrete surface is higher than that of the sandpaper surfaces.

In applications involving rubber in contact with very rough surfaces, such as road or concrete surfaces or sandpaper, the rubber only interacts with the roughness above the average surface plane. To account for this, one should use the top power spectra in theoretical calculations, which assume randomly rough surfaces. The top power spectra are obtained by replacing the roughness below the average plane with roughness that has the same statistical properties as that above the average plane. Similarly, one can define the bottom power spectra. For sandpaper surfaces, all three power spectra (top, bottom, and full) are nearly identical, but for the concrete surface, which is smoother above the average plane than below, this is not the case.

To illustrate this, Fig. \ref{1x.2h.concrete.sandpaper180.80.eps} shows the measured surface height $h(x)$ as a function of distance $x$ along a straight $25 \ {\rm mm}$ track on the concrete surface (red) and on the sandpaper P80 (blue) and P180 (green) surfaces. Note that the upper part of the height profile on the concrete surface is smoother than the lower part. The horizontal dashed lines indicate the average surface plane $\langle h \rangle = 0$, and the data for the P80 and P180 surfaces are shifted downward by $0.2 \ {\rm mm}$ and $0.4 \ {\rm mm}$, respectively.

Fig. \ref{1logq.2logC.concrete.T.B.Full.eps} shows the full (red), bottom (blue), and top (green) surface roughness power spectrum as a function of the wavenumber for the concrete surface. In the calculations in Sec. 5 and 6, we used the top power spectrum; using the full power spectrum would result in a wear rate approximately 20 times higher than reported in Sec. 5, and of similar magnitude to that observed for the sandpaper surfaces (see below). This remarkable result shows the importance of using the top power spectrum for wear and friction calculations.

Fig. \ref{1logq.2logC.P80.P100.P180.eps} shows the full surface roughness power spectrum as a function of the wavenumber for sandpaper surfaces P80 (red), P100 (green), and P180 (blue). The top power spectrum, which is used in the calculations below, is nearly identical to the full power spectra, as expected from the topography images in Fig. \ref{1x.2h.concrete.sandpaper180.80.eps}.

\begin{figure}
\includegraphics[width=0.47\textwidth,angle=0.0]{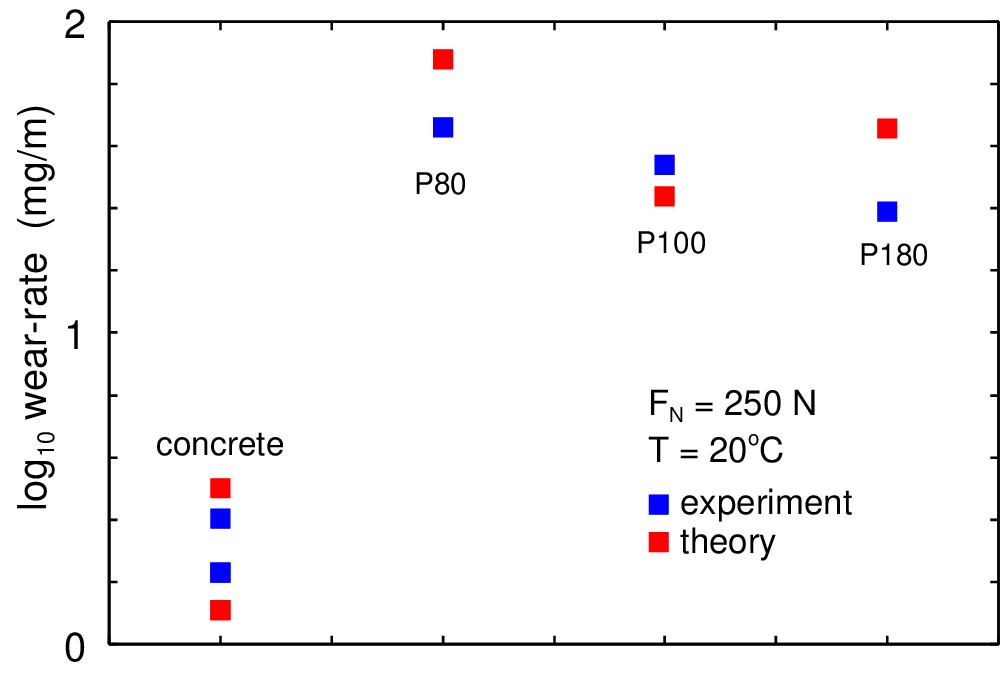}
\caption{\label{1surfaceused.2logWearRate.eps}
The logarithm of the wear rate (in mg/m) for rubber blocks sliding on concrete, and on the sandpaper surfaces
P80, P100 and P180. The blue squares are the experimental results and the red squares
the theory prediction.
The load on concrete is $F_{\rm N} = 250 \ {\rm N}$ and the results for the wear rate for the sandpaper surfaces
was scaled by $250/118$ so they can be compared to the theory and the results for concrete.
}
\end{figure}

In Fig. \ref{1surfaceused.2logWearRate.eps}, we show the logarithm of the wear rate (in mg/m) for rubber blocks 
sliding on the concrete surface and on the sandpaper surfaces P80, P100, and P180. 
The blue squares are the experimental results, and the red squares show the theoretical predictions 
(at a nominal contact pressure of $0.12 \ {\rm MPa}$). We have used the friction coefficients
$\mu = 0.9$ (concrete), $1.19$ (P80), $1.06$ (P100) and $1.11$ (P180). The load on the rubber block 
is $F_{\rm N} = 250 \ {\rm N}$. The experiments on sandpaper were performed at a 
load of $118 \ {\rm N}$ but we assume that the wear rate is proportional to the normal force
and scaled the wear rate by $250/118$ for comparison with the theory and the experiments on the concrete surface.

In a very interesting paper Tanaka et. al.\cite{interest} have studied the friction and wear
for a rubber block sliding at low speed ($1 \ {\rm mm/s}$) on a grinding wheel. 
The wear rate they observed (about $2 \ {\rm mm^3/m}$, when scaled to the same load as we used)
is very similar to what we observed on concrete.
Grinding wheels are produced in a mold by squeezing together hard particles and a 
binder (e.g., a resin) under high pressure between two smooth, parallel surfaces. 
This process results in a surface where the particle tops are nearly at the same height, 
creating what we refer to as smooth roughness in this paper, despite the large rms-roughness.

We consider the agreement between theory and experiment in Fig. \ref{1surfaceused.2logWearRate.eps} remarkable. 
We note that the results depend sensitively on the power spectra, which are somewhat noisy as 
they only involve averaging over three line scans, each $25 \ {\rm mm}$ long.

\vskip 0.3cm
{\bf 8 On the magnitude of $\Delta x$}

The theory presented in Sec. 5 depends on the crack tip displacement $\Delta x$ induced by the interaction with a single wear-asperity. The displacement $\Delta x$ determines the number of contacts $N_{\rm cont} \approx r_0/\Delta x$ needed to remove a wear particle. $\Delta x$ depends on the (oscillatory) driving force or tearing energy $\gamma$. 

For natural rubber with carbon black filler at room temperature, when $\gamma \approx 200 \ {\rm J/m^2}$, the displacement $\Delta x$ is only $\approx 1 \ {\rm nm}$ per oscillation cycle 
(see Ref. \cite{tear1}). If this result holds also at the length scale of wear particles, for crack energy of $\gamma \approx 200 \ {\rm J/m^2}$, one would get that $N_{\rm cont} \approx (10 \ {\rm \mu m})/(1 \ {\rm nm}) = 10^4$ contacts would be needed to remove a rubber particle of $\sim 10 \ \mu m$ size, with even more required for larger particles. We found that using a $\Delta x$ approximately 50 times larger (about $50 \ {\rm nm}$) gives wear rates in agreement with experimental results. 
Thus, around $N_{\rm cont} \approx 100$ contacts with wear-asperities are needed to remove a rubber particle of $\sim 10 \ {\rm \mu m}$ size.

The larger $\Delta x$ required to match the observed wear rate may be due to a reduction in the viscoelastic contribution to the crack propagation energy, as expected for the small-scale systems relevant to rubber wear. Additionally, strain crystallization, which reduces $\Delta x$, may also play a role. While strain crystallization has been observed in macroscopic experiments for natural rubber, it might behave differently at the microscale.

The region where strain crystallization occurs in macroscopic experiments is very large, extending $\sim 1.6 \ {\rm mm}$ from the crack tip, as reported in Ref. \cite{straincrystal}. It has been observed even when the applied tensile strain is below the critical strain required for crack growth initiation. It is also important to note that strain crystallization takes time. In macroscopic experiments, approximately $0.1 \ {\rm s}$ is needed for strain crystallization to occur \cite{time}. In the wear process, the rubber-road asperity interaction time is of the order $\Delta t \approx r_0/v$, and for the highest sliding speed in our study, $v = 1 \ {\rm cm/s}$, the interaction time is approximately $10^{-3} \ {\rm s}$ (using $r_0 = 10 \ {\rm \mu m}$). Thus, at this sliding speed, no strain crystallization is expected. If strain crystallization is absent, natural rubber may exhibit similar $\Delta x$ as observed for other types of rubber, e.g., unfilled 
styrene-butadiene (SB) rubber has $\Delta x \approx 20 \ {\rm nm}$ at $\gamma = 200 \ {\rm J/m^2}$.

The energy per unit crack surface area required to propagate a crack at constant velocity in a viscoelastic solid such as rubber is usually written as
$$G(v,T) = G_0 (v,T) [1+f(v,T)]$$
where $G_0(v,T)$ is the energy to break the strong covalent bonds at the crack tip, and $[1+f(v,T)]$ accounts for viscoelastic energy dissipation in front of the crack tip. The factor $G_0$ is usually assumed to be independent of crack tip velocity and equal to the critical tearing energy $\gamma_0$ for an oscillatory strain (where $\Delta x \rightarrow 0$ as $\gamma \rightarrow \gamma_0$ in the absence of stress corrosion).
However, a detailed study shows that $G_0 (v,T) = g(v,T) G_{00}$ also depends on crack tip speed and temperature, though this dependency is usually much weaker than that of the viscoelastic factor $[1+f(v,T)]$ (see Ref. \cite{JCP}).

At high crack tip speeds, viscoelastic energy dissipation occurs farther from the crack tip. However, for a very small (circular) crack, the stress field in the vicinity of the crack tip follows the singular form $\sim r^{-1/2}$ only out to distances on the order of the crack size, limiting the volume of the region where viscoelastic dissipation occurs. This results in a lower $[1+f(v,T)]$ factor than would be expected for an infinitely long crack in an infinitely extended solid.

\begin{figure}
\includegraphics[width=0.4\textwidth,angle=0.0]{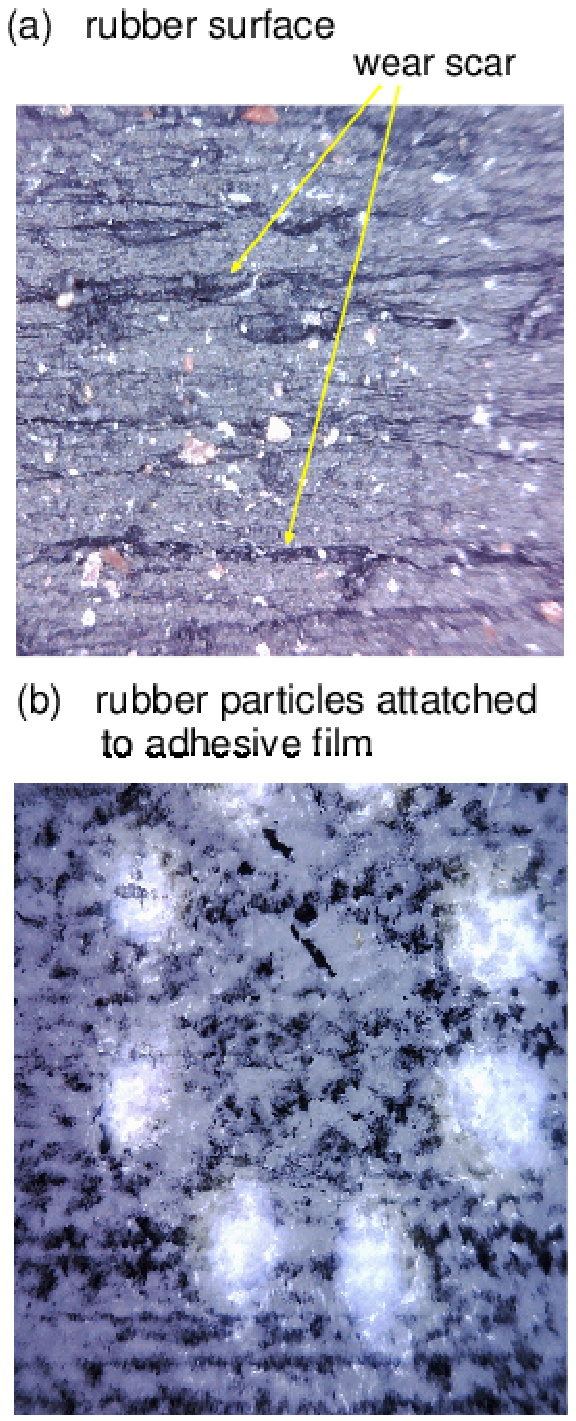}
\caption{
(a) Optical picture of the surface of a rubber block
after the block had been sliding on the sandpaper P100 surface at the nominal contact pressure
$\sigma_0 \approx 0.6 \ {\rm MPa}$.
(b) Optical pictures of an adhesive film which was squeezed against the surface of a rubber block
in (a). Note the linear wear tracks in (a) and the ordering of the wear particles along lines in (b).} 
\label{WearScar1.ps}
\end{figure}

Fig. \ref{WearScar1.ps}(a) shows an optical image of the surface of a rubber block after it had been sliding on the sandpaper P100 surface at a nominal contact pressure of $\sigma_0 \approx 0.6 \ {\rm MPa}$. Fig. \ref{WearScar1.ps}(b) shows rubber particles on an adhesive film that was pressed against the surface in (a). Note the alignment of the wear particles along the lines. This could indicate that the wear particles are not formed randomly on the surface but rather in rows where particularly sharp asperities from the sandpaper cut into the rubber. An alternative explanation is that the wear particles form randomly on the rubber surface after multiple interactions with road asperities but become aligned in lines due to ``combing" by the road asperities.

Fig. \ref{WearScar1.ps}(a) shows what appear to be a few linear wear tracks on the same rubber surface. If these tracks resulted from road asperities that cut and removed rubber particles in single asperity contacts (such that $N_{\rm cont}=1$), it would imply local stresses so high that the energy release rate corresponds to a tearing energy $\gamma$ close to the ultimate tear strength $\gamma_{\rm c}$. Assuming that $\gamma_{\rm c}$ for cracks at the micrometer length scale is similar to that at the macrometer scale (where $\gamma_{\rm c} \approx 10^4-10^5 \ {\rm J/m^2}$), this value is much higher than the elastic energy release rate predicted by the theory.

An alternative explanation for the wear tracks could be that cracks on the rubber surface form and 
grow through interactions with multiple road asperities, with the final removal occurring from only 
a ``few" of the highest and sharpest asperities. In this case, one would expect to see wear tracks on the rubber surface. 
If this explenation is correct one expect a run-in time period with reduced wear until the 
distribution of surface creacks reach their final steady-state configuration.
More studies are needed to resolve this problem.


\begin{figure}
\includegraphics[width=0.47\textwidth,angle=0.0]{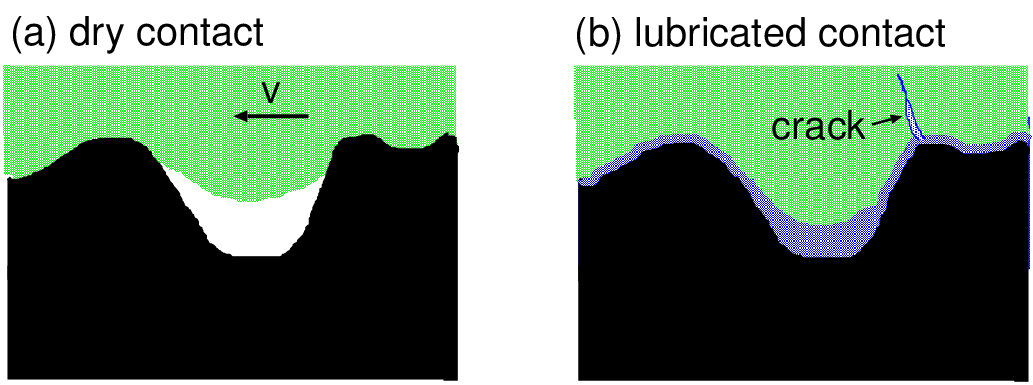}
\caption{
In a fluid, the frictional shear stress in the area of real contact may be reduced
which could allow the rubber to penetrate deeper into a road cavity. This will
increase the tearing stress acting on the rubber in the cavity, which could increase
the rubber wear.} 
\label{ReduceSherStress.eps}
\end{figure}

\vskip 0.3cm
{\bf 9 Discussion}

We have found that the wear rate increases linearly with the load for a carbon-filled natural rubber (NR) compound sliding on a concrete surface at low sliding speeds ($v \le 1 \ {\rm cm/s}$). Similar results were obtained in an earlier study for an SB rubber compound (unpublished). We conclude that, at low sliding speeds and nominal contact pressures relevant to many practical applications, rubber wear is proportional to the normal force. This is the result one intuitively expected, as the real contact area is proportional to the normal force (for sufficiently low normal forces), and in a large enough system, as the normal force increases, new contacts form in such a way that the probability distribution of contact sizes remains unchanged.

However, this finding disagrees with most earlier studies, which show a wear rate that increases faster than linear with the normal force. We attribute this discrepancy to frictional heating, which becomes significant at sliding speeds above $\sim 1 \ {\rm cm/s}$. Frictional heating shifts the viscoelastic modulus $E(\omega )$ master curve to higher frequencies, reducing the viscoelastic contribution to the tearing energy and thereby increasing the wear rate.

We have also found that for our NR compound on concrete, the rubber wear rate in water (and soapy water) is much lower than in the dry state, despite only a small change in the friction coefficient. This result differs qualitatively from an earlier (unpublished) study for an SB compound under the same conditions, where the wear rate was approximately five times higher in water than in the dry state, despite a similarly small change in the friction coefficient. One explanation for the increased wear in water could be a reduction in the frictional shear stress\cite{cut}, which would allow the rubber to penetrate deeper into roughness cavities (see Fig. \ref{ReduceSherStress.eps}), 
resulting in stronger tearing forces acting on the rubber. This explanation is consistent with the observation that it is easier to cut rubber with a sharp knife blade when the contact is lubricated by water or soapy water; this could be due to a reduction in adhesion and frictional shear stress between the blade and the rubber, which concentrates more of the normal force at the sharp knife edge. However, we cannot explain why the same effect does not increase the wear rate in water for the NR compound used in the present study.

We have found that the wear rate is independent of the sliding speed for the range of sliding speeds used in this study (from $1 \ {\rm \mu m/s}$ to $1 \ {\rm cm/s}$). This was also observed in an earlier study for an SB rubber compound. This result is surprising, as one would expect that if a crack is subjected to a constant driving force, where the elastic energy release rate exceeds the value $\gamma_{\rm 0}$ required for crack propagation, the crack tip will move at a constant velocity depending on the energy release rate. Since contact time is inversely proportional to sliding speed, one would anticipate increased wear at lower sliding speeds; however, this is not observed. 

As previously mentioned, this behavior is only possible if either (a) the crack tip motion is of the stick-slip-stick type, where, after a crack tip displacement $\Delta x$, the elastic energy release rate drops enough to prevent further crack motion until another road asperity makes contact with the crack, or (b) a rubber particle is removed with each new asperity contact (with the number of new asperity contacts depending only on sliding distance). We believe the first scenario (a) may be correct, unless the roughness is very sharp, in which, case (b) may hold. In either case, however, the macroscopic relationship between $\gamma$ and $\Delta x$ may not hold at the asperity contact level, as it might predict a crack tip displacement that is too low.

The theory (and experimental results) presented above can be used to estimate the contribution of wear to rubber friction. The energy dissipated during sliding a distance $L$ is $\mu F_{\rm N} L$. 
If $A_{\rm w}$ denotes the total surface area of the formed wear particles, then the energy required to form the wear particles is $A_{\rm w} \gamma$, where $\gamma$ is an effective (or average) fracture energy. 
Thus, the fraction of the total dissipated energy needed for the wear process is $\eta = (A_{\rm w}/L) \gamma / (\mu F_{\rm N})$.

For our sliding block, with $F_{\rm N} = 250 \ {\rm N}$, $\mu \approx 1$, $\gamma \approx 500 \ {\rm J/m^2}$, and $L = 1 \ {\rm m}$, the model calculations give a fracture area of $A_{\rm w} \approx 4 \times 10^{-5} \ {\rm m^2}$. A similar estimate can be obtained directly from the experiments: the observed wear volume after sliding $L = 1 \ {\rm m}$ is $\Delta V \approx 1 \ {\rm mm^3}$. An average wear particle has a size of $r_0 \approx 20 \ {\rm \mu m}$. The number of wear particles is $N = \Delta V / (2 \pi r_0^3 / 3)$, and the fracture surface area is $A_{\rm w} = N \times 2 \pi r_0^2$, giving $A_{\rm w} \approx 3 \Delta V / r_0 \approx 10^{-4}$. 

This results in $\eta \approx 10^{-4}$, indicating that the contribution of wear to the friction coefficient 
is negligible.

\vskip 0.3cm
{\bf 10 Crack tip shielding}

A complete theory of rubber wear must accurately predict the sizes of the wear particles produced. This depends on the variation of the crack-tip displacement $\Delta x$ with the tearing (or crack) energy $\gamma$, as discussed in Sec. 4. It is also influenced by crack-tip shielding. If a ``long" crack exists (which may ultimately result in the removal of a large particle), it is unlikely that smaller cracks will propagate or extend in its vicinity. This is because a long crack reduces the stress in its surrounding area, thereby decreasing the driving force for smaller cracks.

Molecular dynamics simulations reveal that two nearby contact junctions interact elastically when the distance between them is on the order of the junction diameter \cite{Moli}. These elastic interactions result in crack shielding, so that during wear particle formation, not all cracks can fully develop as they are unloaded by nearby propagating cracks, leading to the formation of larger wear particles.

\begin{figure}
\includegraphics[width=0.47\textwidth,angle=0.0]{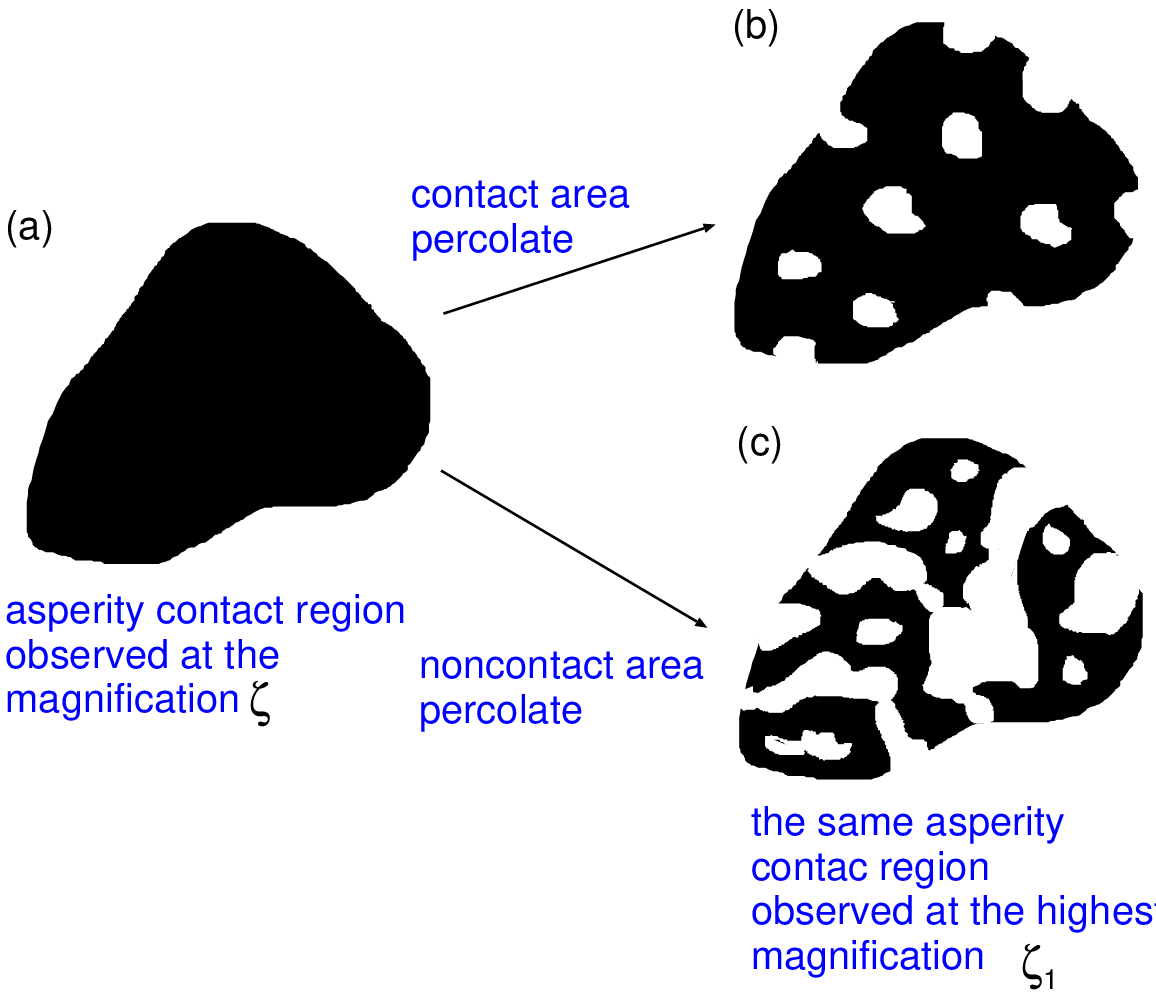}
\caption{\label{AasperityAfterPercolation.eps}
Consider the contact between two elastic solids at the magnification $\zeta$. We define an asperity contact region as compact if no non-contact regions can be observed within it. Generally, an asperity contact region that appears compact at magnification $\zeta$ becomes non-compact at the highest magnification $\zeta_1$, as shown in (b), often consisting of several separated regions as in (c). The magnification $\zeta^*$ is defined such that (on average) a contact region compact at magnification $\zeta^*$ reaches the percolation threshold at the highest magnification. We can determine $\zeta^*$ using $A(\zeta_1)/A(\zeta^*) \approx 0.42$. Here, we use the fact that for a randomly rough surface, the non-contact area percolates when $A(\zeta) /A_0 \approx 0.42$.}

\end{figure}

\vskip 0.3cm
{\bf 11 On the size of contact regions}

Wear particles are removed in asperity contact regions, and the sizes of these contact regions are of central importance. In the wear particle theory presented in Sec. 5, the sizes were determined by the magnification. Here, we present a different approach where the effective size of the contact regions is determined by the stress-stress correlation function. The treatment below is inspired by the studies of M\"user et. al. on the
size of contact regions and its relation to the stress-stress correlation function\cite{Mu1,Mu2}.

We begin with a qualitative discussion on the nature of asperity contact regions. At low magnification $\zeta$, relatively large and compact contact regions can be observed, as illustrated in Fig. \ref{AasperityAfterPercolation.eps}(a). An asperity contact region is considered compact if no non-contact regions can be observed within it. Generally, an asperity contact region that appears compact at magnification $\zeta$ becomes non-compact at the highest magnification $\zeta_1$, as shown in (b), often breaking up into several separated regions as in (c).

The magnification $\zeta^*$ is defined such that (on average) a contact region that appears compact at magnification $\zeta^*$ reaches the percolation threshold at the highest magnification. We can determine $\zeta^*$ using $A(\zeta_1)/A(\zeta^*) \approx 0.42$. Here, we use the fact that for a randomly rough surface, the non-contact area percolates when $A(\zeta) /A_0 \approx 0.42$. We refer to the contact regions observed at magnification $\zeta^*$ as the macroasperity contact regions.

At low magnification (but not so low that the contact area percolates) and at sufficiently low nominal contact pressure, the asperity contact regions are compact and well separated. In this limit, the size of the asperity contact regions is well-defined. However, as illustrated in Fig. \ref{AasperityAfterPercolation.eps}, at high magnification it is less clear how to define the size of the asperity contact regions. One approach is to use the stress-stress correlation function.

Consider the stress-stress correlation function
$$g({\bf x},{\bf x}') = 
\langle \sigma ({\bf x}) \sigma ({\bf x}')\rangle - \langle \sigma ({\bf x}) \rangle 
\langle \sigma ({\bf x}') \rangle.$$
The applied stress is defined as $\sigma_0 = \langle \sigma ({\bf x}) \rangle$, and from now on, we assume the applied stress is subtracted from $\sigma ({\bf x})$, so that $\langle \sigma ({\bf x}) \rangle = 0$.

We define the effective asperity contact radius $r_0$ using the condition $g(r_0) = \alpha g(0)$, where $\alpha < 1$.

Note that $g(r)$ depends on the range of surface roughness included in the calculation. If $q_0$ and $q_1$ are the smallest and largest roughness wavenumbers used in calculating $g(r)$, we include the roughness with wavenumbers $q < \zeta q_0$, where $1 < \zeta < q_1/q_0$. Thus, $g(r) = g(r, \zeta)$ will depend on the magnification $\zeta$.

We consider surfaces with roughness that have statistical properties that are translation invariant. In these cases, $g({\bf x},{\bf x}')$ will depend only on ${\bf x}-{\bf x}'$. By writing
$$\sigma ({\bf x}) = \int d^2q \ \sigma ({\bf q}) e^{i {\bf q}\cdot {\bf x}},$$
we get
$$\langle \sigma ({\bf x}) \sigma ({\bf x}')\rangle = \int d^2q \, d^2q' \ \langle \sigma ({\bf q}) \sigma ({\bf q}') \rangle e^{i ({\bf q}\cdot {\bf x}+{\bf q}'\cdot {\bf x}')}.$$

Since $g$ depends only on ${\bf x}-{\bf x}'$, it remains unchanged if we replace ${\bf x} \rightarrow {\bf x}+{\bf y}$ and ${\bf x}' \rightarrow {\bf x}' + {\bf y}$. If we integrate over ${\bf y}$ and use 
$${1\over A_0}\int d^2y \ e^{i({\bf q}+{\bf q}')\cdot {\bf y}} = {(2\pi )^2\over A_0} \delta ({\bf q}+{\bf q}'),$$ 
we get 
$$\langle \sigma ({\bf x}) \sigma ({\bf x}')\rangle = {(2\pi )^2\over A_0} \int d^2q \ \langle \sigma ({\bf q}) \sigma (-{\bf q}) \rangle e^{i {\bf q} \cdot ({\bf x}-{\bf x}')}.$$

For surfaces with isotropic roughness $g({\bf x})$ depends only on $r=|{\bf x}|$ and we denote it by
$g(r)$. For this case we have shown in Ref. \cite{stresscorrelation}
$$\langle \sigma ({\bf q}) \sigma (-{\bf q})\rangle = {A_0\over (4\pi)^2} (E^*)^2  q^2 C(q) W(q)$$
where $E^* = E/(1-\nu^2)$ and where
$$W(q) = P(q) [\gamma + (1-\gamma)P^2(q)]$$
where
$$P(q)={\rm erf}\left ({\sigma_0 \over 2 \surd G}\right )$$
$$G= {\pi \over 4} ( E^*)^2 \int_{q_0}^q dq \ q^3 C(q)$$
Using () and () we get
$$\langle \sigma ({\bf x}) \sigma ({\bf x}')\rangle = {1\over 4} (E^*)^2  \int d^2q  \ q^2 C(q) W(q) e^{i {\bf q} \cdot ({\bf x}-{\bf x}')}$$
Denoting $|{\bf x}-{\bf x}')| = r$ we get
$$\langle \sigma ({\bf x}) \sigma ({\bf x}')\rangle = {1\over 4} (E^*)^2 2 \pi \int d q  \ q^3 C(q) W(q) J_0(qr)$$
This integral is of the form
$$F=\int_{q_0}^{q_1} dq \ f(q) J_0 (qr)$$
where $f(q)$ is a relatively slowly varying function of $q$. The $q$-integral may be over $\sim 7$ decades
in wavenumber and the distance $r$ can take values from 
$\sim 1/q_1$ to $\sim 1/q_0$ which may be 7 decades in length scale, e.g., for road
surfaces from ${\rm cm}$ to ${\rm nm}$. In Appendix D we show how this integral can be evaluated.

Fig. \ref{AasperityAfterPercolation.eps} shows the radius of the asperity contact area as a 
function of the cut-off wavenumber for $\alpha = 0.25$, $0.5$ and $0.75$. 
Note that for $\alpha = 0.25$ and $0.5$, the radius changes very weakly with increasing wavenumber (or magnification $\zeta = q/q_0$)
for $q>q^*$, or $\zeta > \zeta^*$, where $\zeta^*$ is 
defined by $A(\zeta_1)/A(\zeta^*) = 0.42$, where $\zeta_1 = q_1/q_0$
is the highest magnification. For $\alpha = 0.5$ the radius $r_0 \approx 0.1 \ {\rm mm}$, 
at the magnification $\zeta = \pi/q_0r_0 \approx 2\times 10^4$, where $A_{\rm wear}(r_0)$ is maximal.
Note that, remarkably, the magnification where $A_{\rm wear}(r_0(\zeta))$ is maximal is almost the same as $\zeta^*$.

\begin{figure}
\includegraphics[width=0.47\textwidth,angle=0.0]{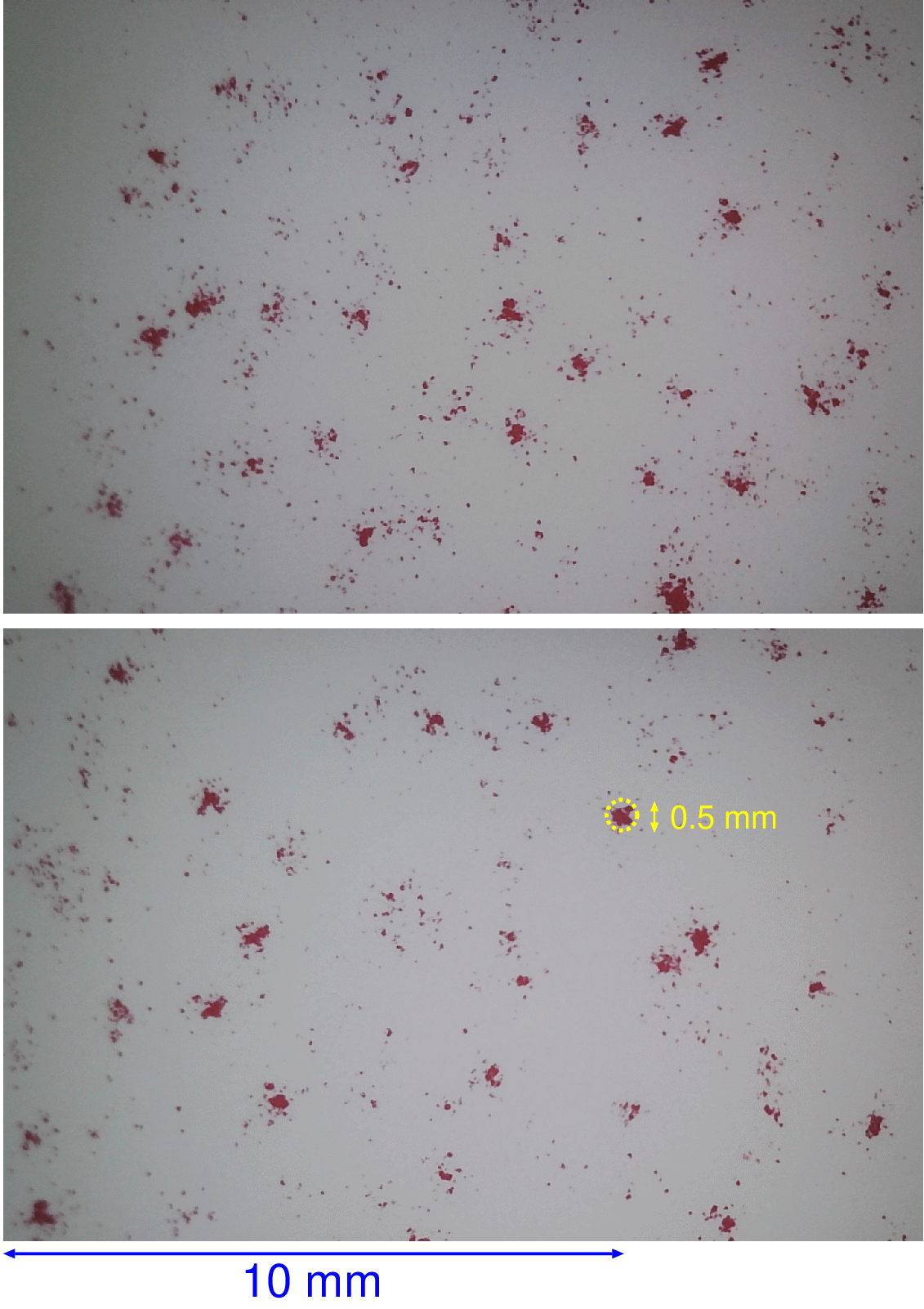}
\caption{
Picture of the contact regions between a rubber block and the concrete surface
at two different surface areas of the concrete block. The nominal
contact pressure $\sigma_0=0.12 \ {\rm MPa}$ is the same as in the rubber wear
experiments. The picture is obtained with a pressure-sensitive film 
(Fujifilm, Super Low Pressure, $0.5-2.5 \ {\rm MPa}$ pressure range) 
between the rubber block and the concrete surface. The dashed circle has the diameter $0.5 \ {\rm mm}$
and most macroasperity contact regions are slightly smaller (typical diameters $0.2-0.4 \ {\rm mm}$.} 
\label{TwoPicConcreteBig.eps}
\end{figure}

We have studied the contact between the rubber block and the concrete surface using a pressure-sensitive film (Fujifilm, Super Low-Pressure film, $0.5-2.5 \ {\rm MPa}$ pressure range, $\lambda = 30 \ {\rm \mu m}$ lateral resolution \cite{Fuji}). Fig. \ref{TwoPicConcreteBig.eps} shows the optical image of the contact region on two different surface areas of the concrete block after $5 \ {\rm s}$ of contact time. The nominal contact pressure $\sigma_0 = 0.12 \ {\rm MPa}$ is the same as that used in the rubber wear experiments. 

\begin{figure}
\includegraphics[width=0.47\textwidth,angle=0.0]{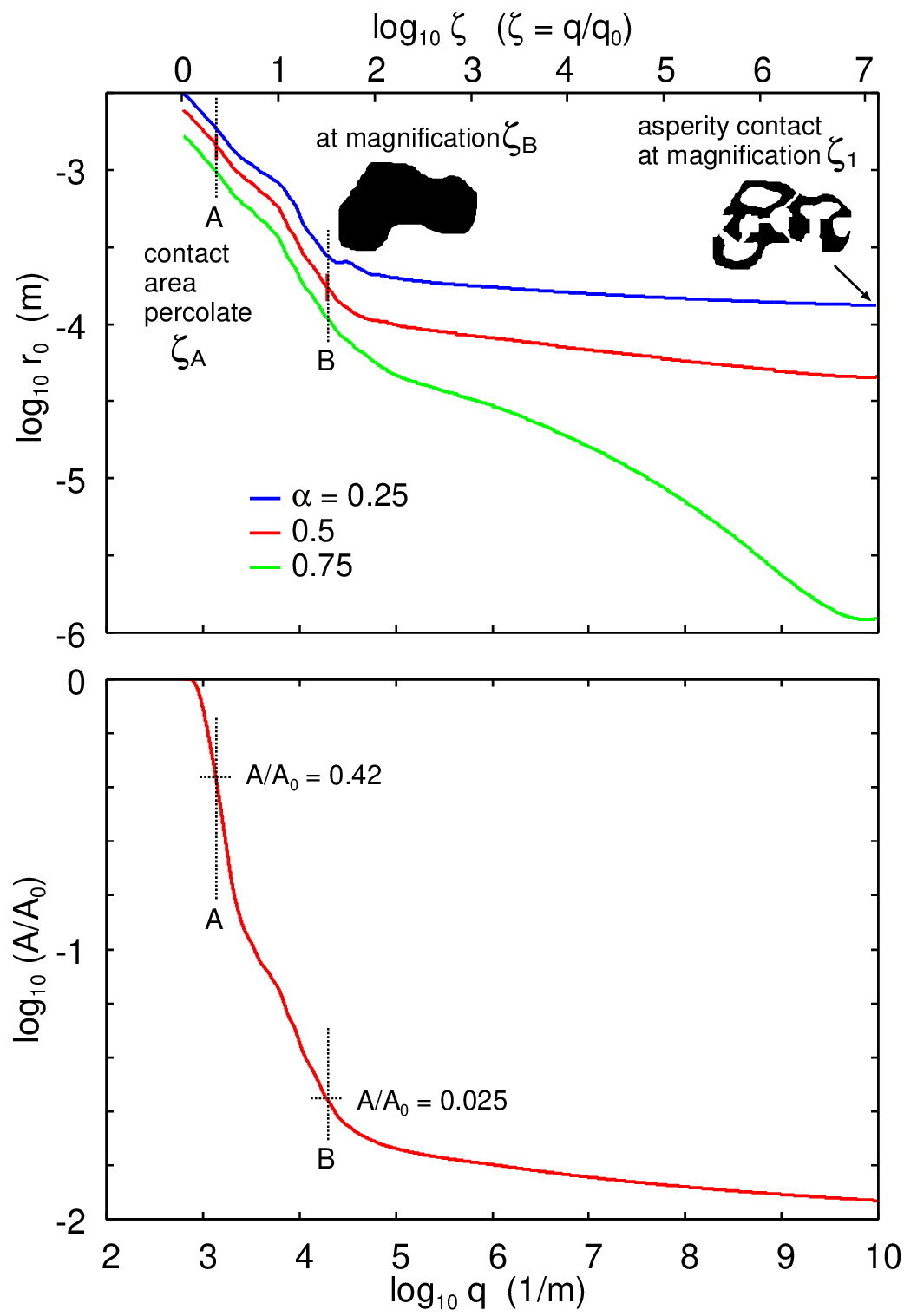}
\caption{\label{1logq.2logRadius.half=0.0.0.75.0.25.eps}
(a) The asperity contact radius $r_0$ as a function of the 
magnification (upper scale) (log-log scale) for $\alpha = 0.25$ (blue line),
$0.5$ (red) and $0.75$ (green). 
The lower scale is the logarithm of the wavenumber of the largest wavenumber roughness 
component included in the calculation ($q=\zeta q_0$).
At the magnification $\zeta_{\rm A}$ the non-contact area percolates.
At the magnification $\zeta_{\rm B}=\zeta^*$ 
an asperity contact region which is compact at this magnification,
will be at the percolation threshold when observed 
at the highest magnification. We have $A(\zeta_{\rm A})/A_0 \approx 0.42$ and
$A(\zeta_1)/A(\zeta^*) \approx 0.42$. 
(b) The relative contact area $A/A_0$ as a function of the cut-off wavenumber (log-log scale).
We have used the rubber elastic modulus $E=10 \ {\rm MPa}$ and Poisson ratio
$\nu = 0.5$. 
}
\end{figure}

\begin{figure}
\includegraphics[width=0.47\textwidth,angle=0.0]{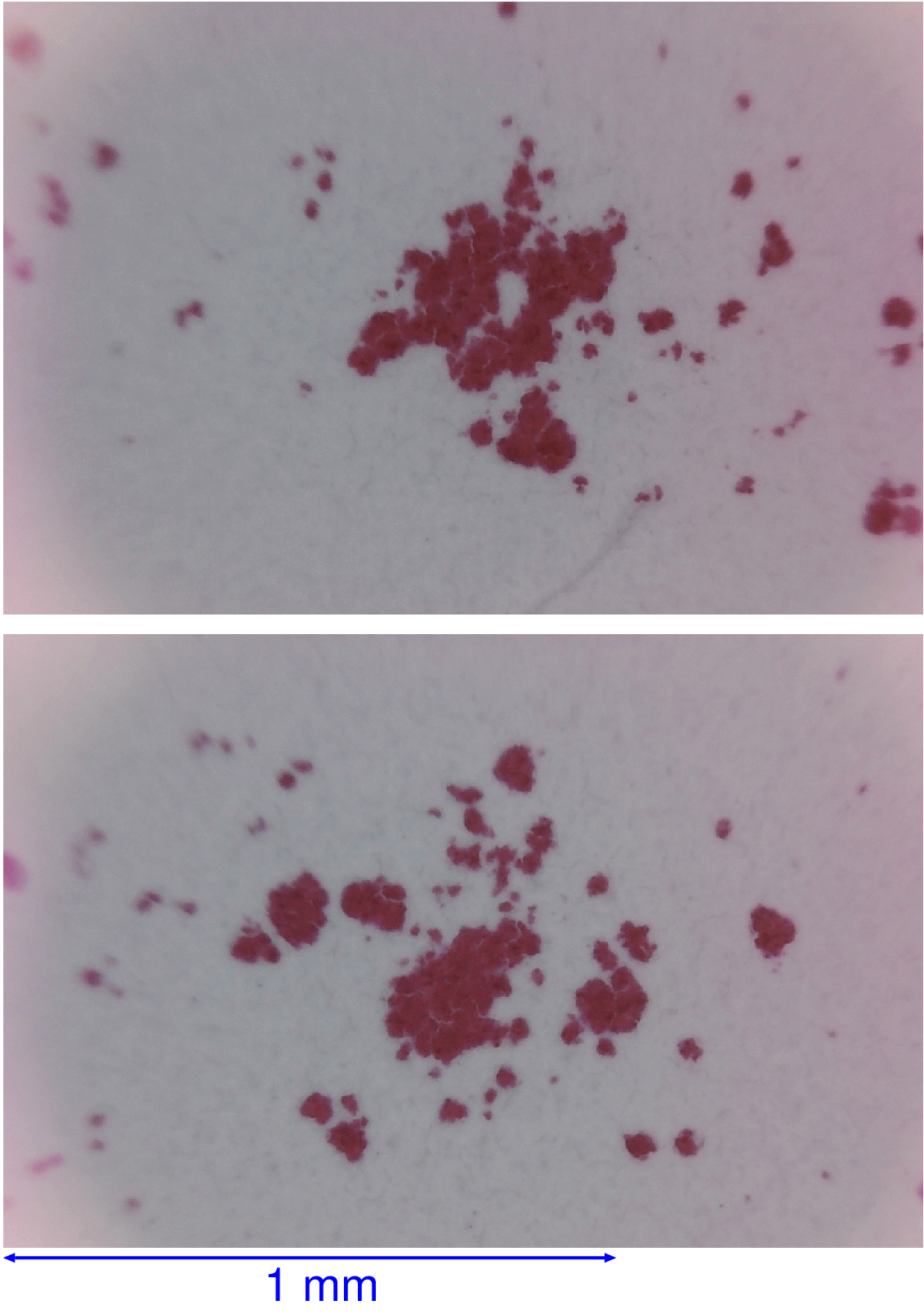}
\caption{
Magnified views of two contact regions from Fig. \ref{TwoPicConcreteBig.eps}.
The linear size of the (macroasperity) contact regions are $2r_0 \sim 0.2-0.4 \ {\rm mm}$.
} 
\label{TwoPicConcreteSmall.eps}
\end{figure}

Fig. \ref{TwoPicConcreteSmall.eps} shows magnified views of two contact regions from Fig. \ref{TwoPicConcreteBig.eps}. The effective (or average) diameter of the (macroasperity) contact regions is $2r_0 \sim 0.2-0.4 \ {\rm mm}$, which agrees with the predicted size of the macroasperity contact regions (indicated by line B in Fig. \ref{1logq.2logRadius.half=0.0.0.75.0.25.eps}), which equals $2r_0 \approx 0.34 \ {\rm mm}$ if $\alpha = 0.5$. 

Estimation of the relative contact area gives $A/A_0 \approx 0.015$, which is close to the calculated result $A/A_0 \approx 0.017$ obtained from Fig. \ref{1logq.2logRadius.half=0.0.0.75.0.25.eps}(b) for $q=\pi/\lambda$, where $\lambda = 30 \ {\rm \mu m}$ is the lateral resolution. In the calculation, we assumed a modulus of $E=10 \ {\rm MPa}$; however, this is not accurately known as it depends on the strain in the asperity contact regions and the contact time due to viscoelastic relaxation.

\vskip 0.3cm
{\bf 12 Summary and conclusion}

We have presented an experimental and theoretical study of the wear rate for a rubber block sliding on concrete surfaces. The most important results are:

(1) The measured wear rate is independent of the nominal contact pressure (or load) as it varies from $\sigma_0 = 0.12$ to $0.43 \ {\rm MPa}$. It is also independent of sliding speeds from $1 \ {\rm \mu m/s}$ to $1 \ {\rm cm/s}$.

(2) The wear rate in water and soapy water is too small to be detected using our approach, which measures the weight of the rubber block before and after a given sliding action.

(3) The independence of the wear rate on the sliding speed indicates that the crack tips do not move continuously when exposed to the stress field induced by a road asperity. Instead, during sliding, stress builds up at the crack tip, but the crack does not move until the stored elastic energy reaches a critical value. After this point, crack tip motion occurs at a speed unrelated to the sliding speed and hence unrelated to the rubber-road asperity interaction time.

This observation was also made in an earlier (unpublished) study involving two SB rubber compounds. This result shows that what matters for wear is not the contact time but rather the sliding distance. Although slower speeds result in longer contact times between the rubber and substrate asperities, this does not result in greater crack propagation distances as one might anticipate. This independence arises because cracks propagate in discrete steps rather than continuously; specifically, they follow a stick-slip pattern. During sliding, strain energy builds up at the crack tip until it reaches a critical threshold, after which the crack advances incrementally by a displacement $\Delta x$. This process results in a 
wear rate governed primarily by the cumulative number of asperity contacts per unit sliding distance, rather than by contact time. Consequently, the crack tip motion occurs at a speed unrelated to the relative speed of rubber-asperity contact, leading to a wear rate that remains constant across the tested range of sliding speeds.

(4) A theory for rubber wear is developed based on the probability distribution of stress acting at the interface, as determined by the Persson contact mechanics theory. Rubber wear particles can form at an asperity contact region when the stored elastic energy due to shear stress exceeds the energy required for bond breaking and viscoelastic dissipation needed to form the wear particle.

(5) A theory and numerical results for the size of the rubber-concrete contact regions as a function of magnification were presented. The theory predicts macroasperity contact regions with a radius of $r \approx 0.1 \ {\rm mm}$, which is nearly the same as the theoretical predictions for the radius of the wear particles that gives the biggest contribution to the wear volume.

The predicted probability distribution for the size of the wear particles has a maximum at $\approx 20 \ {\rm \mu m}$, which appears consistent with photos of wear particles removed from the rubber surface using adhesive film.

The wear rates we calculate depend rather sensitively on most of the parameters used in the theory. This may explain why, when repeating a wear experiment days or months later, we usually obtain somewhat different wear rates even if the experiments are conducted under nominally identical conditions. For example, we measured the wear rate for the same rubber compound on nominally identical concrete blocks and found different values at different times (separated by a few months): $2.5 \ {\rm mg/m}$, $1.7 \ {\rm mg/m}$, and $1.0 \ {\rm mg/m}$. These variations may reflect small changes in the concrete block surface topography (some blocks were bought at different times and may be from different batches), 
or fluctuations in humidity or temperature, or aging of the rubber samples (which were kept in a refrigerator at $+4 \ {\rm C}$ when not in use).

To improve the accuracy of wear calculations using the theory presented above, it is necessary to study how the relationship between the tearing energy $\gamma$ and the crack tip displacement $\Delta x$ is modified at short-length scales. We suggest that this could be done by examining the propagation of small cracks (length $\sim 10 \ {\rm \mu m}$) in thin rubber sheets (thickness $\sim 10 \ {\rm \mu m}$). It is also important to determine the parameter $\beta$ in Eq. (3) by comparing theoretical wear predictions with experimental data. Although $\beta$ is expected to be approximately 1, even a small variation in this parameter could significantly impact the wear rate.

\vskip 0.3cm
{\bf Acknowledgments:}
We thank A. Almqvist, 
M. M\"user and R. Stocek
for useful comments on the text. 
We thank M. M\"user for suggesting to define the size of 
contact regions using the stress-stress correlation function.
This work was supported by the Strategic Priority Research 
Program of the Chinese Academy of Sciences, Grant No. XDB0470200.

\vskip 0.5cm
{\bf Appendix A: Stress probability distribution}

Consider the stress at the interface between an elastic half-space 
and a rigid surface with random roughness. When we study the interface at the magnification $\zeta$
we only observe surface roughness with wavenumber $q<\zeta q_0$ (or wavelength $\lambda > \lambda_0/\zeta$
with $q=2\pi/\lambda$ and $q_0=2\pi/\lambda_0$), 
where $q_0$ is the wavenumber of the most long wavelength roughness included in the study.
Let $P(\sigma, \zeta)$ denote the probability distribution of stress which depends on the magnification.
The expression (5) for $P(\sigma, \zeta)$ can be derived from the Persson contact mechanics 
theory as is described elsewhere, and briefly reviewed here. 

Assuming complete contact between an elastic half-space and a rigid surface with random roughness
it can be shown that\cite{JCPP}
$${\partial P \over \partial \sigma} = D(\zeta) {\partial^2 P \over \partial \sigma^2}\eqno(A1)$$
where the ``diffusivity'' $D$ depends on the elastic properties of the solids and on the surface roughness
power spectrum $C(q)$ (note: $q=\zeta q_0$):
$$D = {\pi \over 4} \left ({E\over 1-\nu^2}\right )^2 q_0 q^3 C(q)\eqno(A2)$$
It is now assumed that (A1) holds {\it locally} also when partial contact occurs at the interface.
At the lowest ``engineering'' magnification $\zeta=1$ we do not observe any roughness, e.g., a road surface
appear flat without roughness, as assumed in most engineering studies of the 
contact between a tire and a road surface.
At this magnification, friction can be neglected if uniform stress is applied 
at the upper surface of a block, the same uniform stress acting at the interface,
which gives the (initial) boundary condition:
$$P(\sigma,1) = \delta (\sigma-\sigma_0)\eqno(A3)$$
One can show that for any magnification, as $\sigma \rightarrow 0$ the $P\rightarrow 0$ giving the boundary condition
$$P(0,\zeta)=0\eqno(A4)$$
In addition, there can be no infinite stress at the interface giving
$$P(\infty, \zeta)=0\eqno(A5)$$
Solving the stress diffusion equation (A1) using the boundary conditions (A4) and (A5) and the
(initial) condition (A3) gives (5).

\vskip 0.5cm
{\bf Appendix B: Rubber effective modulus}

The characteristic deformation frequency when a rubber block 
slides in contact with a road asperity
is $\omega \approx v/r_0$, where $\sim r_0$ is the 
linear size of the contact region. In the present case 
the contact radius 
$\approx 0.01 \ {\rm mm}$ and sliding speeds $v=1-10^4 \ {\rm \mu m/s}$ giving
deformation frequencies between $0.1 \ {\rm s}^{-1}$ and $1000  \ {\rm s}^{-1}$. In this frequency range
the low strain modulus change from $\approx 27$ to $36 \ {\rm MPa}$.
For the large strain relevant here we now show that 
the effective relevant modulus is $E \approx 10-20 \ {\rm MPa}$. 

The stress-strain relation for filled rubbers is strongly non-linear. 
We define the effective (secant) modulus $E$ as the ratio between the (physical) stress and 
the strain $E=\sigma /\epsilon$. Here $\sigma = F/A$, where $F$ is the elongation force and $A$
the rubber block cross-section of the rubber strip which depends on the strain, $A=A_0 (1+\epsilon)$,
where $A_0$ is the cross-section of the rubber strip before elongation.
The strain $\epsilon $ is defined in the usual way $\epsilon = (L-L_0)/L_0$ where $L$ and $L_0$
are the length of the rubber strip before and after applying the force $F$.
We need this secant modulus for the typical strain prevailing in the 
rubber-road asperity contact regions. 

\begin{figure}
\includegraphics[width=0.47\textwidth,angle=0.0]{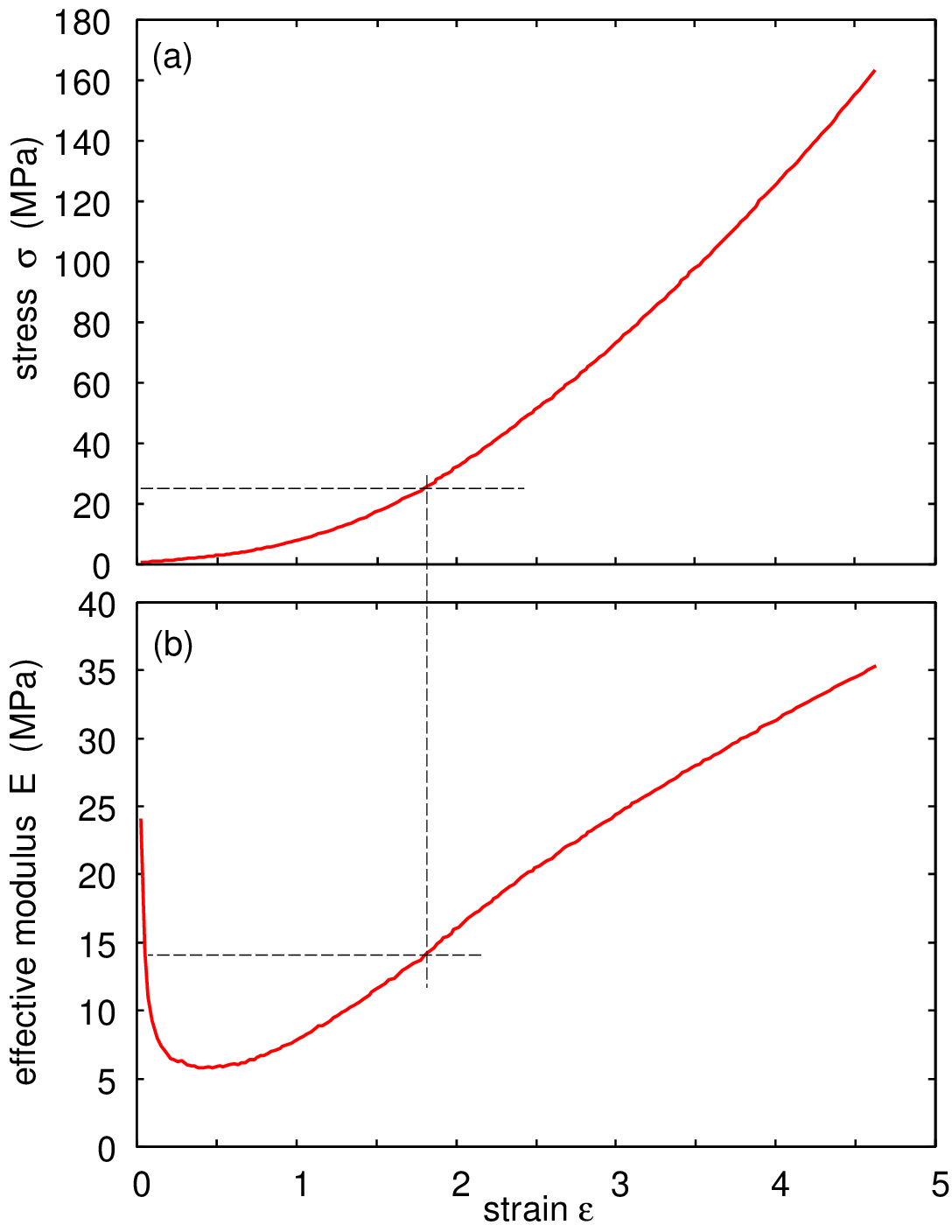}
\caption{
(a) The strain-stress relation for the natural rubber used in the
present study. The stress is the true (or physical) stress defined by $\sigma = F/A$ where $F$ is the elongation force and $A$
the rubber block cross section which depends on the strain. The strain is defined as usual $\epsilon = (L-L_0)/L_0$.
(b) The effective (secant) modulus $E = \sigma/\epsilon$. The measurement was performed by elongation a strip of rubber
at the strain rate $\dot \epsilon = 0.3 \ {\rm s}^{-1}$.}
\label{1strain.2stress.eps}
\end{figure}

The stress in the asperity contact regions can be estimated using (4) which with
$E^* \approx 15 \ {\rm MPa}$, $\gamma \approx 200 \ {\rm J/m^2}$ and $r_0 \approx 10 \ {\rm \mu m}$
gives $\sigma \approx 25 \ {\rm MPa}$. 
Fig. \ref{1strain.2stress.eps}(a) 
shows the stress-strain relation for the elongation of a strip of the natural rubber used in the
present study. The stress $\sigma \approx 25 \ {\rm MPa}$ correspond to the strain $\epsilon \approx 1.8$. 
The effective (secant) modulus $E$ for this strain is shown in Fig. \ref{1strain.2stress.eps}(b) 
and is about $E=14 \ {\rm MPa}$.

\begin{figure}
\includegraphics[width=0.47\textwidth,angle=0.0]{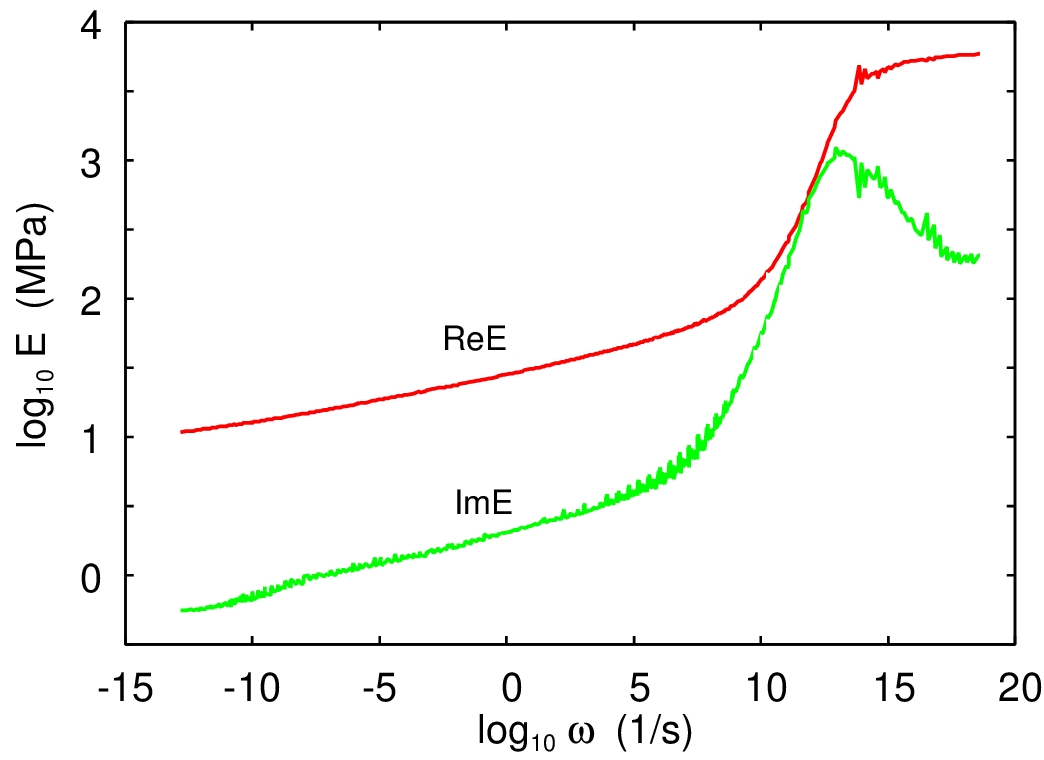}
\caption{
The dependency of the low strain ($\epsilon = 0.0004$) viscoelastic modulus 
	on the frequency  $\omega$ for $T=20^\circ {\rm C}$ (log-log scale).}
\label{1logOmega.2logE.bus.eps}
\end{figure}

\vskip 0.5cm
{\bf Appendix C: Surface roughness power spectrum}

The most important quantity of a rough surface is the surface roughness power spectrum\cite{Challange,Review}.
The two-dimensional (2D) surface roughness power spectrum $C({\bf q})$,
which enters in the Persson contact mechanics theory,
can be obtained from the height profile $z=h(x,y)$ measured over a square 
surface unit. However, for surfaces with
roughness having isotropic statistical properties, the 2D power spectrum 
can be calculated from the 1D power spectrum obtained
from a line-scan $z=h(x)$.

The 2D power spectrum is defined by\cite{JCPP,Preview}
$$C({\bf q}) = {1\over (2\pi )^2} \int d^2x \ \langle h({\bf x})h({\bf 0}) \rangle 
e^{-i{\bf q}\cdot {\bf x}}\eqno(B1)$$
If we write the surface
profile $z=h(x,y)$, given on a two-dimensional (2D) square surface area, as the sum (or integral) of plane waves
$$h({\bf x}) = \int d^2q \ h({\bf q}) e^{i{\bf q}\cdot {\bf x}}, \eqno(2)$$
then the 2D power spectrum can also be written as
$$C({\bf q}) = {(2 \pi )^2\over A_0} |h({\bf q})|^2, \eqno(3)$$
where $A_0$ is the surface area. For surfaces with isotropic statistical properties $C({\bf q})$ depends only on
the magnitude $q=|{\bf q}|$ of the wavevector ${\bf q}$. We can write $q=2 \pi /\lambda$, where $\lambda$ is the wavelength
of a surface roughness component.

Many surfaces, including the concrete surfaces studied here (see Fig. \ref{1x.2h.concrete.sandpaper180.80.eps}), display a 
non-symmetric height distribution (i.e., no symmetry as $h \rightarrow -h$). For such 
surfaces it is interesting to study the top power spectra $C_{\rm T}$ defined by\cite{Preview}
$$C_{\rm T} = {1\over (2\pi )^2} {A_0 \over A_{\rm T}} \int d^2x \ 
\langle h_{\rm T}({\bf x})h_{\rm T}({\bf 0}) \rangle 
e^{-i{\bf q}\cdot {\bf x}}\eqno(B4)$$
where $h_{\rm T} ({\bf x}) = h({\bf x})$ for $h > 0$ and zero otherwise, and where $A_T/A_0$ is the fraction of the
total (projected) area where $h>0$.
In a similar way one can define the bottom power spectrum $C_{\rm B}$ using $h_{\rm B}({\bf x}) = h({\bf x})$ 
for $h < 0$ and zero otherwise. 
The physical interpretation of $C_{\rm T}$ is that it is the power spectrum of a surface where the roughness below
the average plane is replaced by roughness with the same statistical properties as above the average plane. This is the relevant power spectra to use in theory calculations which assume random roughness that ``looks the same'' above and below
the average surface plane.

The wear experiments have been performed on concrete and sandpaper surfaces.
The concrete blocks (concrete pavers) were obtained in a large number from a ``Do-It-Yourself'' shop.
In most cases, every new wear experiment was done on a 
new concrete block. We have used these concrete blocks in most of our earlier friction studies.
They are very stable (no or negligible concrete wear), and concrete blocks obtained from the same batch 
have all the same nominal surface roughness. 
For each surface we measured at least 3 tracks at different locations, each $25 \ {\rm mm}$ long.

\begin{figure}[!ht]
\includegraphics[width=0.9\columnwidth]{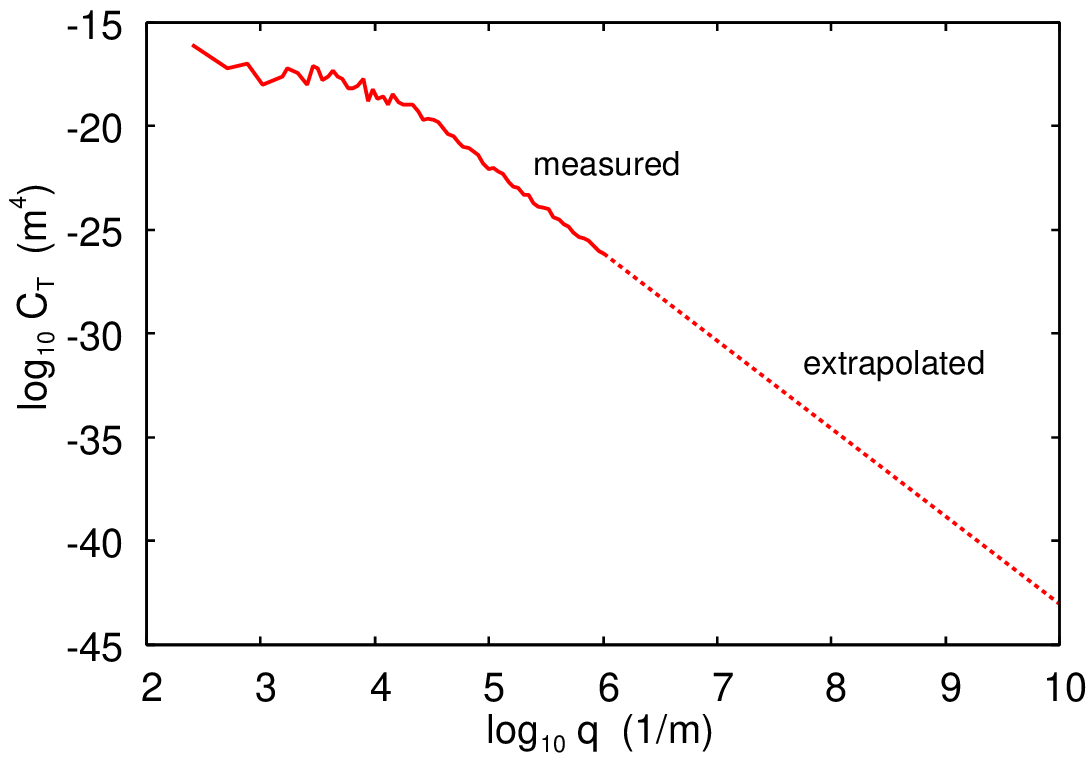}
\caption{\label{1logq.2logC.concrete.T.eps}
        The 2D top surface roughness power spectrum obtained from line scans on 
the concrete surface (solid line) and the linearly
        extrapolated power spectrum (dashed line). The linear extrapolated region 
corresponds to the Hurst exponent $H\approx 1$.
}
\end{figure}

We have measured the surface topography using
a Mitutoyo Portable Surface Roughness Measurement Surftest SJ-410 equipped with a diamond tip 
having a radius of curvature $R=1 \ {\rm \mu m}$, and with the tip-substrate repulsive 
force $F_N = 0.75 \ {\rm mN}$. The step length (pixel) is $0.5 \ {\rm \mu m}$, 
the scan length $L=25 \ {\rm mm}$ and the tip speed  $v=50 \ {\rm \mu m/s}$. 
The top power spectra shown in Fig. \ref{1logq.2logC.concrete.T.eps} 
(solid line) was obtained by averaging over three measurements.
The dashed line is the linearly
extrapolated power spectrum. The linear extrapolated region corresponds to the Hurst exponent $H\approx 1$.
The power spectra of the sandpaper surfaces were shown in Sec. 6 and were extrapolated linearly 
to the same large wavenumber cut-off $q_1$ as in Fig. \ref{1logq.2logC.concrete.T.eps}.

\vskip 0.5cm
{\bf Appendix D: Contact radius integral}

We need to calculate
$$F=\int_{q_0}^{q_1} dq \ f(q) J_0 (qr)\eqno(D1)$$
where $f(q)$ is a relatively slowly varying function of $q$. The $q$-integral may be over 7 decades
in wavenumber and the distance $r$ can take values from $\sim 1/q_1$ to $\sim 1/q_0$ so over 7 decades in length scale. We write
$$q= q_0 e^\mu$$
Using that $ dq = q_0 e^\mu d\mu$ we get
$$F=\int_0^{\mu_0} d\mu \ q f(q) J_0 (qr)$$
where $\mu_0 = {\rm ln}(q_1/q_0)$. We write
$\mu=n \Delta$, where $n=1,2,.., N$, $\Delta << 1$, $N= \mu_0/\Delta$,
so that
$$F\approx \sum_{n=0}^{N}  q_n f(q_n) Q_n\eqno(D2)$$
where
$$Q_n = \int_{n \Delta}^{(n+1) \Delta} d\mu \ J_0 (q r)$$
Write $qr=x$ or $x=q_0 r e^\mu$ giving $dx= x d\mu$ and
$$Q_n = \int_{x_n}^{x_{n+1}} dx  \ {1\over x} J_0 (x)\eqno(D3)$$
$$x_{n+1} = q_0 r e^{(n+1)\Delta} = x_n e^\Delta \approx x_n (1+\Delta)$$
If $x_{n+1}-x_n \approx x_n \Delta$ is small enough (say $<100$) 
we can do the integral in (D3) directly.
For large $x_{n+1}-x_n$ we write 
$$Q_n = \int_{x_n}^{x_n^*} dx  \ {1\over x} J_0 (x)
+\int_{x_{n+1}^*}^{x_{n+1}} dx  \ {1\over x} J_0 (x)$$
$$+\int_{x_n^*}^{{x_{n+1}^*}} dx  \ {1\over x} J_0 (x)\eqno(D4)$$
where $x_n^*$ is the smallest number of the form $2 \pi i+\pi/4$ which is larger than $x_n$, and where
where $x_{n+1}^*$ is the largest number of the form $2 \pi j+\pi/4$ which is smaller than $x_{n+1}$. 
The first two integrals in (D4) can be performed by direct integration. The last integral
$$H_n =\int_{x_n^*}^{x_{n+1}^*} dx  \ {1\over x} J_0 (x)$$
can be evaluated as follows. We write
$x =y_k+ y$, where $y_k = 2\pi (i+k)+\pi/4$ and define the integer $N'=j-i$. 
We get   
$$H_n = \sum_k \int_0^{2 \pi} dy  \ {1\over y_k+y} J_0 (y_k+y)$$
where the sum is from $k=1$ to $k=N'$. For large argument of the Bessel function
$$J_0(x) \approx \left ({2\over \pi x} \right )^{1/2} {\rm cos} (x-\pi/4)$$
so we get
$$H_n \approx  \left ({2\over \pi}\right )^{1/2}  
\sum_k \int_0^{2 \pi} dy  \ {1\over (y_k+y)^{3/2}} {\rm cos} (y)$$
Since $y/y_k << 1$ we get
$$H_n \approx  \left ({2\over \pi}\right )^{1/2}  
\sum_k \int_0^{2 \pi} dy  \ \bigg ({1\over (y_k)^{3/2}} - {3\over 2} {y\over (y_k)^{5/2}}$$
$$+{15\over 4}{y^2\over (y_k)^{7/2}}\bigg )  {\rm cos} (y)$$
or
$$H_n \approx  \left ({2\over \pi}\right )^{1/2}  
15 \pi  \sum_k {1 \over (y_k)^{7/2}}$$
$$ \approx \left ({2\over \pi}\right )^{1/2} 15\pi 
\int_1^{N'} dk \ {1 \over [2 \pi(i+k)+\pi/4]^{7/2}} $$
$$= \left ({2\over \pi}\right )^{1/2} 15\pi {2\over 5} \bigg [{1 \over [2 \pi(i+1)+\pi/4]^{5/2}}$$
$$-{1 \over [2 \pi(i+N'+1)+\pi/4]^{5/2}}\bigg ]$$
$$\approx \left ({2\over \pi}\right )^{1/2} 6\pi \left [{1 \over (x_n)^{5/2}}
-{1 \over (x_{n+1})^{5/2}}\right ]$$
Using $x_{n+1} \approx x_n (1+\Delta)$ this gives
$$H_n \approx 
 15 (2\pi)^{1/2}  {\Delta \over (x_n)^{5/2}}\eqno(D5)$$


\begin{thebibliography}{99}
\bibitem{Fuji}
Fuji Pressure Film is a Mylar-based film that contains a layer of microcapsules with a range of sizes
but with typical diameter $\sim 10 \ {\rm \mu m}$.         
The application of pressure upon the film causes the microcapsules to rupture,
producing small color dots and a permanent high-resolution 
image of pressure variation across the contact area. The spatial resolution of the pressure
variations is $\lambda \sim 30 \ {\rm \mu m}$.
\end{thebibliography}
\end{document}